# Heliconical smectic phases formed by achiral molecules

Jordan P Abberley[1] , Ross Killah[1] , Rebecca Walker[1], John MD Storey[1] , Corrie T Imrie[1], Miroslaw Salamonczyk[2,3], Chenhui Zhu[2], Ewa Gorecka[4], Damian Pociecha*[4]

**Affiliations:**

[1] Department of Chemistry, King's College, University of Aberdeen Aberdeen, AB24 3UE, UK

[2] Advanced Light Source, Lawrence Berkeley National Laboratory, Berkeley, 94720 CA, USA.

[3] Department of Physics and Liquid Crystal Institute, Kent State University, Kent, OH 44242, USA

[4] University of Warsaw, Department of Chemistry, Zwirki i Wigury 101, 02-089 Warsaw, Poland.

*Correspondence to: pociu@chem.uw.edu.pl

**Abstract:** A series of asymmetric dimers with an odd number of atoms in the spacer were found to form different types of twisted structures despite being achiral. The formation of a variety of helical structures is accompanied by a gradual freezing of molecular rotation. The tight pitch heliconical nematic ($N_{TB}$) phase and heliconical tilted smectic C ($SmC_{TB}$) phase are formed. In the lowest temperature smectic phase, HexI, the twist is expressed through the formation of hierarchical structure: nano-scale helices and mesoscopic helical filaments. The short pitch helical structure in smectic phases is confirmed by resonant x-ray measurements.

**One Sentence Summary:** The spontaneous formation of helical structures with nano-scale pitch is found to occur in smectic phases consisting of achiral molecules.

The spontaneous formation of chiral structures from effectively achiral molecules is at the heart of intense worldwide research, as the breaking of mirror symmetry is a fundamental issue in chemistry, physics and biology and plays a central role in the origin of biological homochirality. Twisted structures made from chiral building blocks are relatively common in nature, the archetypal examples being DNA or proteins that can exist in helical forms not only in solids but also in the liquid state, whereas examples of achiral molecules that assemble into helical aggregates are much less common, and until recently such helical aggregates were observed only for 3D ordered crystals. This has now changed with the recent discovery of the twist-bend nematic ($N_{TB}$) phase,[1,2] some 35 years after its prediction by Meyer,[3] and which has caused considerable excitement. In the intervening period, Dozov independently predicted the existence of the $N_{TB}$ phase using symmetry arguments.[4] At the root of Dozov's prediction was the assumption that bent molecules have a natural tendency to pack into bent structures, however since uniform bend is not permitted in nature it must be accompanied by other local deformations, namely splay or twist of average molecular axis direction (director). The splay-bend nematic phase is achiral, by contrast, in the $N_{TB}$ phase the director forms a conical left- or right-handed helix. The $N_{TB}$ phase provided the first example of spontaneous chiral symmetry breaking in a fluid with no spatial ordering.[1,2,5-7] The helix in the $N_{TB}$ phase is extremely short, typically around 10 nm (3-4 molecular distances).[5,6,8,9] For the overwhelming majority of twist-bend nematogens, the $N_{TB}$ phase is preceded by a conventional nematic (N) phase with uniform director structure, for which the strong molecular curvature gives rise to small values of the bend elastic constant.[7] In fact, there are just three examples of direct $N_{TB}$–isotropic phase transitions.[10-12] On cooling, the vast majority of $N_{TB}$ phases either crystallize or vitrify, and only rarely is a $N_{TB}$-smectic phase transition observed (see, for example [13,14]). We have yet to establish and understand how the bent molecules will self-assemble into smectic phases if the bend elastic constant becomes anomalously low. Here we show that as in the nematic phase, such achiral molecules also spontaneously form short pitch length helical structures in smectic phases.

The simplicity of the twist-bend nematic phase has significant application potential, and a conventional nematic phase having a small bend elastic constant may itself be utilized in new technological applications such as the electrically controlled selective reflection of light[15] and in an electrically tunable laser.[16]

The molecular curvature, that is essential for the formation of the $N_{TB}$ phase, can be realized using an odd-membered mesogenic dimers. Such dimers are built of two mesogenic moieties linked by a flexible spacer, normally an alkyl chain.[17] The LC behavior of dimers are strongly dependent on the length and parity of the spacer and this is most commonly attributed to how the average shape of the molecule is controlled by conformations of the spacer. Thus, if we consider the spacer to exist in an all-trans conformation then for an even-membered spacer the long axes of the mesogenic groups are essentially parallel and the molecule is linear, whereas for an odd-membered spacer they are inclined at some angle with respect to each other giving a bent molecular shape. Here we report the characterization of the 4-[{[4-({6-[4-(4-cyanophenyl)phenyl]hexyl}oxy) phenyl]methylidene}amino]phenyl-4-alkoxy-benzoates, and use the acronym ***CB6OIBeOn*** to describe them in which *n* represents the number of carbon atoms in the terminal alkyl chain. The molecular structure of the ***CB6OIBeOn*** compounds is shown in Table 1. The hexyloxy spacer was chosen as this provides the molecular curvature required to observe the $N_{TB}$ phase, while increasing the terminal chain length should promote smectic phases.

**Table 1.** Phase transition temperatures (°C) and associated enthalpy changes (in parentheses, kJ mol$^{-1}$) obtained by DSC for the ***CB6OIBeOn*** series. General molecular structure of ***CB6OIBeOn*** is also given.

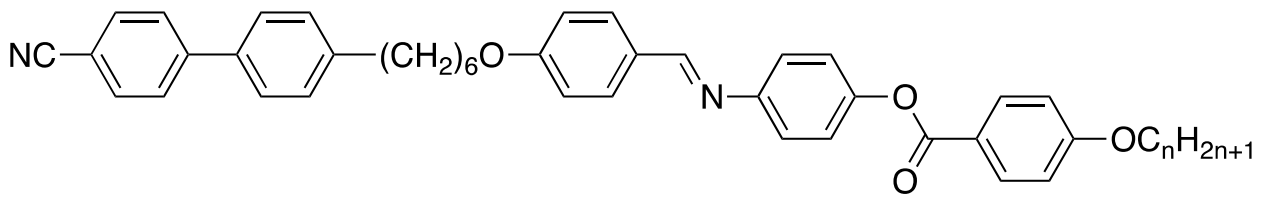

| *n* | m.p. | HexI | | $SmC_{TB}$ | | $SmA_b$ | | SmA | | $N_{TB}$ | | N | | Iso |
|---|---|---|---|---|---|---|---|---|---|---|---|---|---|---|
| 1 | 138.7 (48.3) | | | | | | | | | • | 144.5 [a] | • | 283.5 (2.7) | • |
| 2 | 123.6 (52.2) | • | 56.6 (4.29) | | | | | | | • | 141.0 [a] | • | 279.4 (3.2) | • |
| 3 | 122.3 (36.8) | | | | | | | | | • | 135.7 [a] | • | 267.6 (2.2) | • |
| 4 | 131.1 (47.3) | • | 66.2 (5.1) | | | | | | | • | 132.0 [a] | • | 260.0 (2.1) | • |
| 5 | 127.3 (39.9) | • | 71.5 (5.1) | | | | | | | • | 129.8 [a] | • | 248.8 (1.9) | • |
| 6 | 102.7 (40.4) | • | 73.6 (4.5) | | | | | | | • | 127.0 [a] | • | 243.6 (1.7) | • |
| 7 | 115.4 (54.4) | • | 84.8 (4.6) | • | 104.2 [b] | • | 104.4 [b] (0.07) | • | 122.8 (0.04) | | | • | 236.5 (1.8) | • |
| 8 | 112.2 (49.6) | • | 88.8 (4.3) | • | 100.4 [b] | • | 102.6 [b] (0.06) | • | 143.9 (0.05) | | | • | 231.9 (2.0) | • |
| 9 | 112.0 (50.4) | • | 89.9 (4.6) | | | • | 99.3 (0.07) | • | 158.9 (0.09) | | | • | 224.0 (1.6) | • |
| 10 | 101.4 (56.9) | • | 94.5 (5.6) | | | • | 99.4 (0.02) | • | 172.3 (0.11) | | | • | 222.3 (1.5) | • |

[a] from microscopic observation, [b] $SmA$-$SmA_b$ and $SmA_b$-$SmC_{TB}$ transitions were not resolved in the thermograms, combined enthalpy changes are given, $SmA_b$ phase has been detected only in microscopic observation for *n*=7.

The phase sequences, transition temperatures and associated enthalpy changes are collected in Table 1, and the phase diagram based on calorimetric, X-ray and optical studies is presented in Fig. 1a. It should be noted that studied dimers show a complex polymorphism of liquid crystalline phases. Homologues with short terminal chains, *n*=1-6, exhibit two nematic phases, N and $N_{TB}$, and a lamellar phase below the $N_{TB}$ phase. Both nematic phases give similar XRD patterns, evidencing only short range positional ordering of the molecules (Fig. S1).

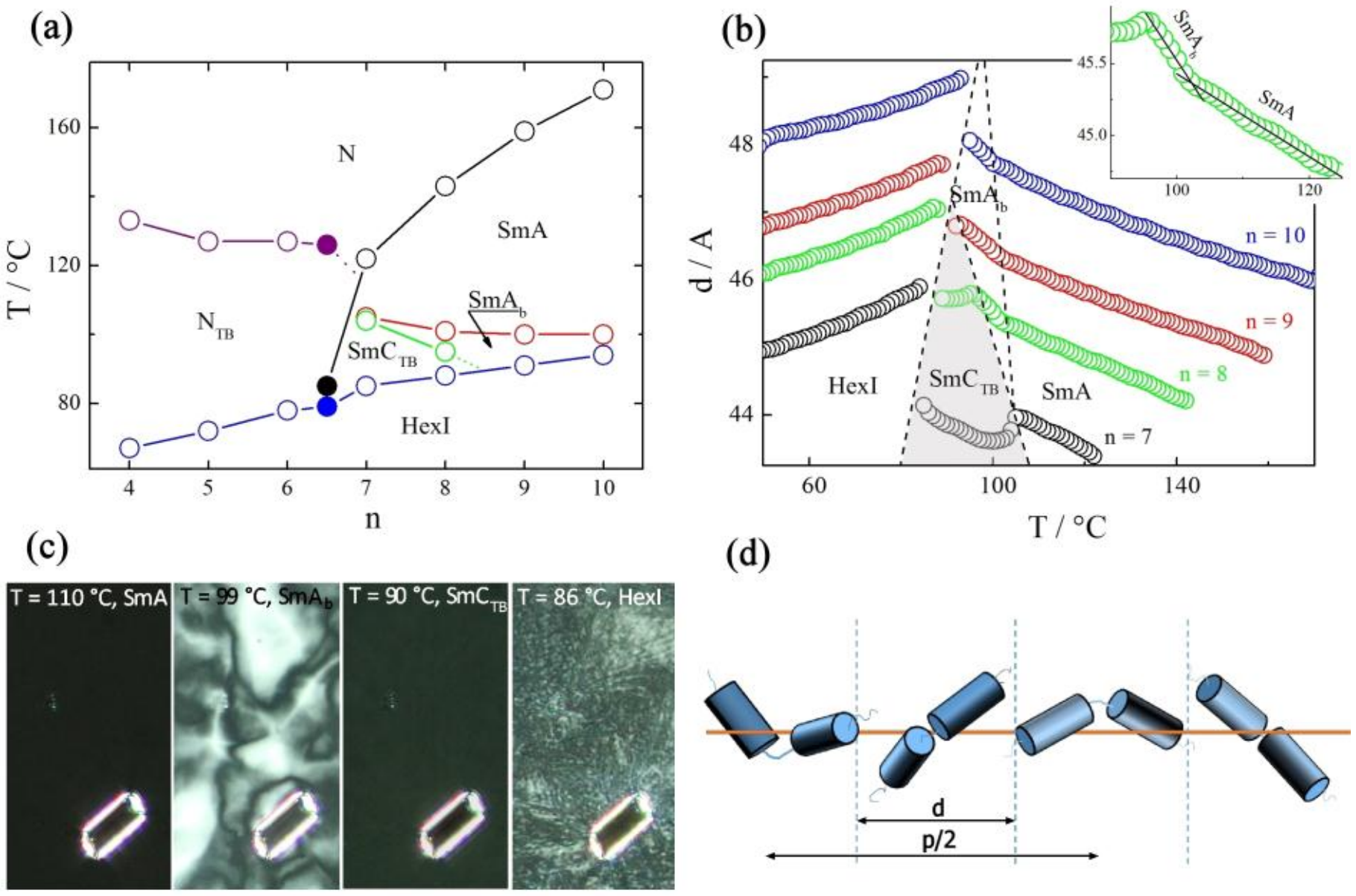

**Fig. 1**. (a) Part of the phase diagram for the achiral ***CB6OIBeOn*** series with *n*=4-10, the filled points show the transition temperatures for a 50 wt. % mixture of homologues n=6 and 7. (b) Smectic layer spacing vs. temperature for ***CB6OIBeOn*** compounds with n=7-10. The d(T) for n=7 and 8 suggests transition from orthogonal to tilted phase. The highlighted area is the heliconical smectic $C_{TB}$ phase. In the inset enlarge fragment of d vs. T for homologue *n*=8 (c) Optical textures of uniaxial SmA, biaxial $SmA_b$, helical $SmC_{TB}$ and HexI phases, observed in an homeotropic cell, between crossed polarizers, for homologue ***CB6OIBeO8***. Glass bead is visible in the picture to mark the same area of the presented textures. Note that the $SmC_{TB}$ phase is uniaxial, which excludes simple synclinic or anticlinic tilted structure. (d) Schematic drawing of the proposed model for $SmC_{TB}$ phase structure in which dimeric molecules form short-pitch helix. The dotted vertical lines denote smectic layer interfaces.

For the N and $N_{TB}$ phases only very weak low angle signals were detected, corresponding to about the full and half molecular length, indicating a local short-range lamellar structure (cybotactic groups). For longer homologues, the $N_{TB}$ phase is extinguished, and instead, up to four lamellar phases were found below the nematic phase. The layer spacing in the smectic phases corresponds approximately to the full molecular length. The lowest temperature lamellar phase is the same for short and long homologues, its X-ray pattern in the high angle range shows a narrowed, split signal. The correlation length of the in-plane positional order, calculated from the width of the high angle diffraction signal, corresponds to several molecular distances, suggesting a hexatic-type smectic phase.[18] The characteristic XRD pattern obtained for an aligned sample, in which one of the high angle signals is seen in an equatorial position with respect to the low angle signals (Fig. S2), shows that the molecules are tilted toward the apex of the local in-plane hexagons, and hence, the phase is assigned as a HexI phase.[19] For all the other smectic phases (the higher temperature smectics), XRD patterns with a single, broad high angle signal were detected, indicating liquid-like positional ordering of the molecules within the smectic layers (Fig. S2). The phase transitions between the smectic phases are clearly visible as changes in the layer thickness measured versus temperature (Fig. 1b). Two of the liquid-like smectic phases, appearing in a sequence below the nematic phase, differ only in their values of the layer thermal expansion coefficient; for example, -0.03 Å $K^{-1}$ and -0.07 Å $K^{-1}$ for ***CB6OIBeO8***. At the phase transition to the lowest temperature liquid-like smectic phase, a continuous decrease of layer spacing is observed, suggesting a transition to a tilted phase (Fig. S3). Finally, a jump in the layer thickness is seen at the transition to the HexI phase. Summarizing, the X-ray studies allowed for the identification of the lamellar phases as two orthogonal smectic A variants, a tilted smectic C and a hexatic smectic I phase.

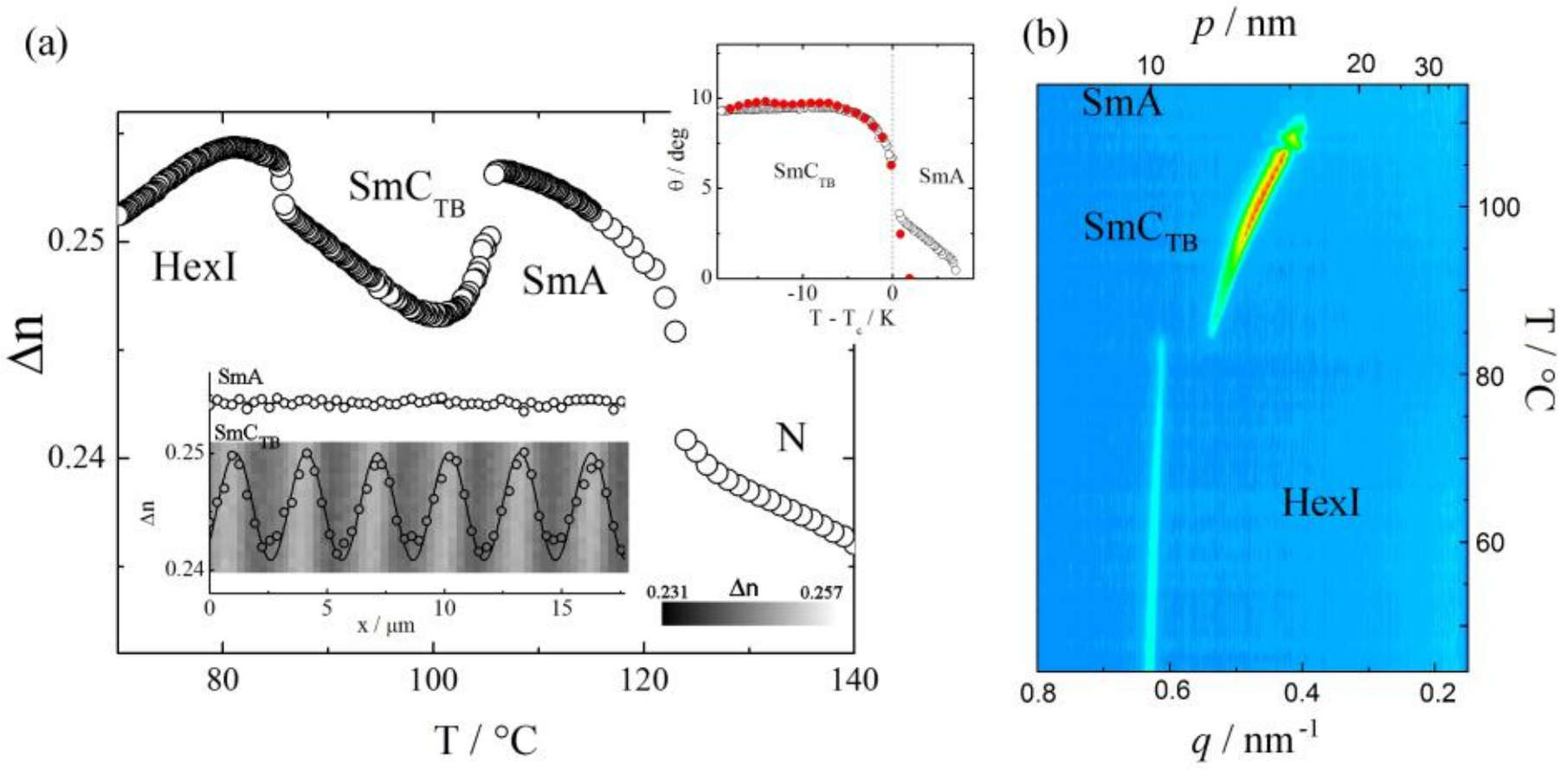

**Fig. 2** (a) Temperature dependence of the optical birefringence, Δn, for homologue ***CB6OIBeO7***, in the lower inset - evolution of birefringence across the stripe area showing a decrease of Δn in $SmC_{TB}$ with respect to the orthogonal SmA phase. The stripes are shown in the background of the inset. In the upper inset the dependence of conical angle θ deduced from birefringence (open circles) and from X-ray measurements (filled, red circles). (b) Temperature evolution of the RSoXS signal for the ***CB6OIBeO7*** compound measured on cooling; while in the HexI phase the signal position is practically temperature independent (it corresponds to ~10 nm), in the $SmC_{TB}$ phase it changes strongly due to the changes of the heliconical pitch length from 11.7 to ~15 nm upon approaching the transition to the SmA phase. Note that the pitch dependence is monotonic, while the smectic layer spacing is non-monotonic function of temperature.

In optical studies the N-$N_{TB}$ phase transition is observed as a change in the microscopic texture. On decreasing the temperature to a few degrees below the transition to the $N_{TB}$ phase, in cells with planar anchoring the uniform texture seen for the nematic phase is replaced by stripes with periodicity equal to the cell thickness (Fig. S4). Interestingly, for the materials studied here a similar texture change is also observed for longer homologues at the transition to the tilted smectic (SmC) phase (Fig. S5). The striped texture is thought to indicate a low bend elastic constant in the $N_{TB}$ phase and the observation of a similar texture in the tilted smectic phase strongly suggests that its bend elastic constant is also very low.[20] The stripes, formed in the $N_{TB}$ or smectic C phase, persist on cooling to the temperature at which the material crystallizes. The optical textures were also studied in cells with homeotropic anchoring and several other observations were noted; the highest temperature smectic phase is optically uniaxial, consistent with a SmA phase assignment;

for lower temperature orthogonal phase a schlieren-like texture is observed, and in the smectic C phase the texture again becomes uniformly black when observed between crossed polarizers (Fig. 1c). The schlieren texture of the orthogonal smectic phase suggests that the phase is optically biaxial ($SmA_b$ phase), in which molecular rotation around the long axis is to some degree frozen, whereas the optical uniaxiality of the tilted smectic C phase, excludes simple synclinic or anticlinic lamellar structures, and suggests an averaging of molecular orientations due to the formation of a short helix – this is a new type of a tilted smectic phase which we term the twist-bend $SmC_{TB}$ (Fig. 1d). The assumption is supported by the temperature dependence of the average optical birefringence, for which a decrease in Δn is observed at the SmA-$SmC_{TB}$ phase transition (Fig. 2a). Also precise determination of birefringence across the stripe area shows the decrease of birefringence in $SmC_{TB}$ phase with respect to the value measured in the smectic A phase (inset Fig. 2a). The conical tilt angle, θ, calculated from the decrease of Δn in the $SmC_{TB}$ phase (for details see SI and [21]) for homologue ***CB6OIBeO7*** saturates at ~10 degree, and the same value of the tilt angle has been deduced from changes of the smectic layer spacing (inset in Fig. 2a). The X-ray studies point towards a weakly first order transition from the orthogonal smectic to tilted smectic for homologue $n$=7 and a continuous transition for homologue $n$=8, with a critical exponent close to 0.4 (Fig. S3). For the $N_{TB}$ phase of homologue ***CB6OIBeO6*** the conical angle,θ, obtained from the decrease of birefringence measured below N-$N_{TB}$ phase transition is ~20 degree. The suggestion that the smectic C phase is heliconical is further supported by results obtained using soft X-ray resonant scattering (RSoXS) at the carbon absorption edge, a technique used previously to probe the helical structure of the $N_{TB}$ phase[9]. On cooling, at the SmA-$SmC_{TB}$ phase transition a sharp, resonant Bragg peak develops, corresponding to the pitch length $L \approx$ 15nm (Fig. 2b). The pitch slightly shortens with decreasing temperature, being incommensurate with layer structure, and corresponds to 3-4 smectic layer distances. The resonant signal has been also detected in the hexatic smectic phase, in this phase the peak position shows only weak temperature dependence (it changes in the range 10-10.2 nm and corresponds to ~2.2 smectic layer distances). The $SmC_{TB}$-HexI transition is first order with a clear discontinuity of the superlayer structure and a small temperature range in which both phases coexist (Fig. S4).

The texture of the HexI phase in cells with homeotropic anchoring is very weakly birefringent (Fig. 1c), and a slight de-crossing of the polarizers reveals the presence of optically active domains (Fig. S6). The HexI phase gives a clear circular dichroism signal around the absorption band of the material (~350 nm), consistent with the chiral nature of the phase (Fig. S7). Therefore it is apparent that in this phase the chiral segregation occurs on a microscopic scale. No CD signal is detected in the $SmC_{TB}$ or $N_{TB}$ phases, so it appears that either the effect is weak or the size of the chiral domains in these phases is smaller than the optical wavelength. The morphology of the sample in the HexI phase was studied by AFM. These studies clearly revealed the presence of untangled helical filaments with an average diameter of ~50 nm (Figs. 3 and S8). For the samples aligned between glass slides with planar anchoring, large areas with uniformly oriented entangled filaments were found. The filaments appear to have uniform twist over micron size areas. The morphology of the sample strongly resembles that observed for the $B_4$ phase, although it should be noted that filaments of $B_4$ phase usually have an internal crystalline structure.[22,23]

No clear optical flexoelectric switching, reported previously for other $N_{TB}$ materials, was found for the studied compounds. Instead, only a weak change in the birefringence and a bi-polar response under high electric fields were detected, except for the HexI phase, which was not sensitive to electric field.

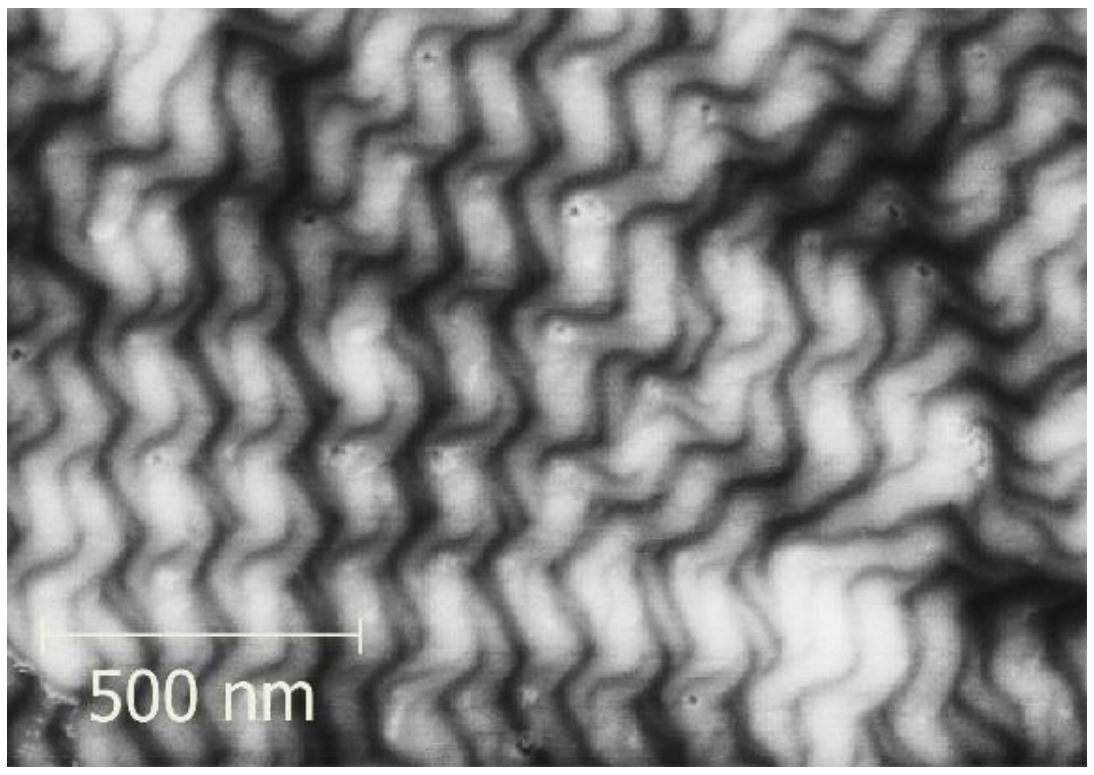

**Fig. 3** AFM image of twisted entangled filaments formed in the temperature range of the HexI phase for homologue ***CB6OIBeO7.***

Summarizing, for the short homologues of the ***CB6OIBeOn*** series the phase transition between the uniform N and helical $N_{TB}$ phases was found, while for longer homologues, as the tendency for the alkyl chains and aromatic moieties to nano-scale separation increases, a complex sequence of smectic phases is observed in which the tilted phases show a helical precession of the molecules around the cone axis, similar to chiral smectic phases. The heliconical phase is preceded by the non-tilted SmA and $SmA_b$ phases. Apparently on lowering the temperature, rotation around the long molecular axis is gradually frozen and as a result the transition from SmA to $SmA_b$ phase is observed. The gradual freezing of rotation is accompanied by a more pronounced layer spacing increase, due to lower interdigitation of molecules between neighbouring layers and a stretching of the terminal chains. The lack of a polar response under an electric field shows that in the $SmA_b$ phase the transverse dipole moments of the molecules are locally compensated. The X-ray studies revealed that upon lowering temperature, for intermediate length homologues, the $SmA_b$ phase undergoes a transition (weakly first order or continuous) to a tilted phase, which surprisingly is optically uniaxial. This optical uniaxiality is inconsistent with a simple synclinic or anticlinic SmC phase structure, and indicates that averaging of molecular orientation must take place through the formation of a helioconical structure. The resonant x-ray data indeed proves the existence of the superstructure, with a pitch corresponding to 3-4 smectic layers. This is the first report showing unambiguously that a short pitch helical structure may be formed spontaneously in smectic phases consisting of achiral molecules. Such short helices were found previously for smectic phases comprising chiral rod-like molecules (SmC$\alpha$ phase) as a result of competing interactions between nearest and next nearest layers. In systems in which next nearest interactions favour antiparallel tilt the system relieves frustration by the formation of a helix that can be as short as a few smectic layers[24]. However molecules studied here are achiral dimers and similarly as for the $N_{TB}$ phase, the helix formation is presumably driven by steric interactions arising from the bent molecular shape. The bent dimers studied have an extremely strong tendency to form twisted structures, this is evidenced by the heliconical arrangement of the molecules in the $N_{TB}$ and $SmC_{TB}$ phases. In the lowest temperature smectic phase, the HexI phase, the twist is expressed through an even more complex, hierarchical assembly with a nano-pitch heliconical structure and mesoscopic helical filaments.

**Acknowledgments:**

MS acknowledges the support of the U.S. National Science Foundation I2CAM International Materials Institute Award, Grant DMR-1411344 and NSF grant DMR-1307674. DP, EG acknowledges the support of the National Science Centre (Poland) under the grant no. 2016/22/A/ST5/00319. RW gratefully acknowledges the Carnegie Trust for the Universities of Scotland for the award of a PhD studentship. The beamline 11.0.1.2 at the Advanced Light Source at the Lawrence Berkeley National Laboratory is supported by the Director of the Office of Science, Office of Basic Energy Sciences, of the U.S. Department of Energy under Contract No. DE-AC02- 05CH11231.

Supplementary Materials for

# Heliconical smectic phases formed by achiral molecules

Jordan P Abberley, Ross Killah, Rebecca Walker, John MD Storey, Corrie T Imrie, Miroslaw Salamonczyk, Chenhui Zhu, Ewa Gorecka, Damian Pociecha.

correspondence to: pociu@chem.uw.edu.pl

## Materials and Methods

Materials

All materials were used as purchased (Sigma Aldrich, Alfa Aesar), unless otherwise stated. Dichloromethane was dried over 3Å molecular sieves prior to use. Solvents were evaporated at approximately 20 mm Hg using a water aspirator pump connected to a Buchi rotary evaporator. Trace solvents were removed from solids in a Thermo Scientific vacuum oven at 1.0 mm Hg and 50 °C. Reactions were monitored using thin layer chromatography (TLC) carried out on aluminium-backed plates with a coating of Merck Kieselgel 60 F254 silica and an appropriate solvent system. Silica gel coated aluminium plates were purchased from Merck KGaA. Spots were visualised using UV light (254 nm) or by oxidation with either an aqueous permanganate dip or iodine.

Synthesis

### Synthesis of 6-bromo-1-[4-(4-bromophenyl)phenyl]hexan-1-one (BrBK5Br).

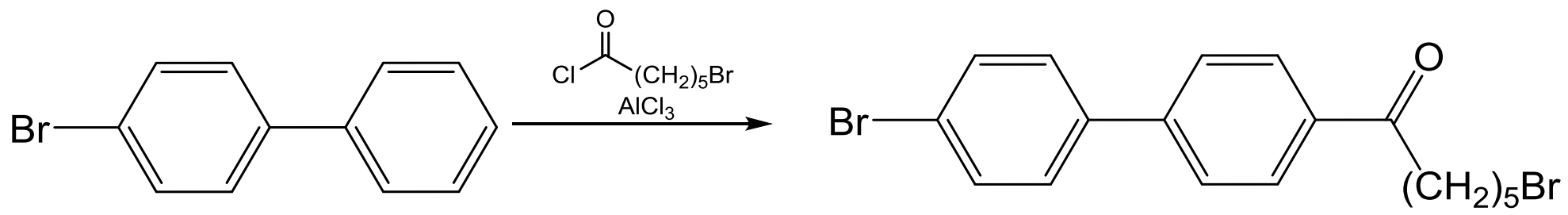

Scheme S1: Synthesis of **BrBK5Br**.

The synthesis of 6-bromo-1-[4-(4-bromophenyl)phenyl]hexan-1-one (**BrBK5Br**) was performed according to Scheme S1 [1]. Significantly greater yields were noted for this reaction when all glassware was wrapped in aluminium foil, and this is presumed to be due to photolytic debromination of the reagent bromobiphenyl *via* a free radical pathway [2]. A stirred mixture of anhydrous aluminium chloride (12.503 g, 9.38 x $10^{-2}$ mol, 1.2 mol. eq.) in dichloromethane (200 ml) was cooled to 0 °C, and to this a solution of 4-bromobiphenyl (18.26g, 7.83 x $10^{-2}$ mol, 1 mol. eq.) and 6-bromohexanoyl chloride (20.00 g, 9.36 x $10^{-2}$ mol, 1.2 mol. eq.) in dichloromethane (50 ml) was added dropwise in the absence of light, over a period of 30 min. After 1 h, the ice was removed and the mixture allowed to reach room temperature with stirring for a further 4 h. The solution was slowly added to a mixture of crushed ice (100 g), distilled water (100 ml) and 20 % hydrochloric acid (100 ml) to destroy the aluminium chloride complex. The organic layer was collected and washed with distilled water (2 x 200 ml), and dried over magnesium sulfate. *In vacuo* removal of the dichloromethane gave an off-white solid which was recrystallised twice from boiling ethanol (200 ml x 2) and dried in a vacuum oven at 50 °C overnight.

***BrBK5Br** - 6-bromo-1-[4-(4-bromophenyl)phenyl]hexan-1-one:*

Yield: 26.52 g, 82.9 %.

Melting point: 81.5 – 82.2 °C.

Infrared $\bar{\nu}$ $cm^{-1}$: 2952, 2866 ($sp_3$ C-H); 1684 (C=O); 1604 (Ar C=C); 807 (p. disubs. benzene C-H), 665 (C-Br).

$^1$H NMR: (300 MHz, Chloroform-d) δ ppm: 1.50 – 1.66 (m, 2 H, -C(O)-$CH_2$-$CH_2$-**$CH_2$**-$CH_2$-$CH_2$-Br), 1.75 - 1.88 (m, 2 H, -C(O)-$CH_2$-**$CH_2$**-$CH_2$-), 1.90 – 2.02 (m, 2 H, -$CH_2$-**$CH_2$**-$CH_2$-Br), 3.05 (t, J=7.3 Hz, 2 H, -C(O)-**$CH_2$**-$CH_2$-), 3.47 (t ,

J=6.7 Hz, 2 H, -$CH_2$-**<u>CH2</u>**-Br), 7.48 – 7.55 (m, 2 H, Ar), 7.59 – 7.70 (m, 4 H, Ar), 8.02 – 8.08 (m, 2 H, Ar).
$^{13}$C NMR: (75 MHz, Chloroform-d) δ ppm: 23.38, 27.89, 32.63, 33.56, 38.35, 122.65, 127.05, 128.72, 128.80, 132.11, 135.99, 138.80, 144.39, 199.40.

**Synthesis of 1-bromo-4-[4-(6-bromohexyl)phenyl]benzene (BrB6Br).**

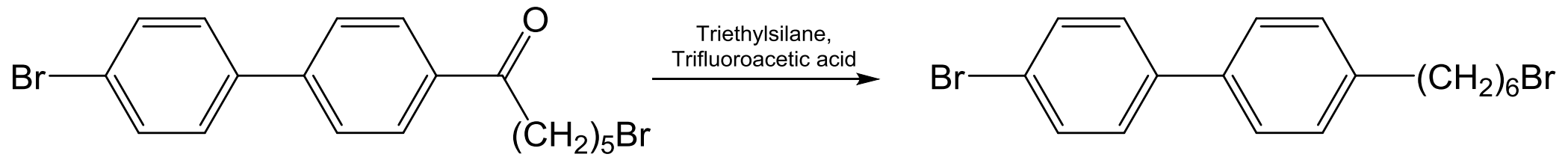

Scheme S2: Synthesis of **BrB65Br**

Scheme The synthesis of 1-bromo-4-[4-(6-bromohexyl)phenyl]benzene (**BrB6Br**) was performed according to Scheme S2, as previously reported by Jansze et al.[3]. Thus, a dry solution of BrBK5Br (26.508 g, 6.47 x $10^{-2}$ mol, 1 mol. eq.) in trifluoroacetic acid (60.0 ml, 3.53 x $10^{-1}$ mol, 5.5 mol. eq.) was cooled to 0 °C, and triethylsilane (23.0 ml, 1.44 x $10^{-1}$ mol, 2.2 mol. eq.) added dropwise over a period of 20 min. After 30 min, dry dichloromethane (40 ml) was added, and left to stir for 1 h under ice and then overnight at room temperature. The product mixture was poured over a mixture of dichloromethane (200 ml) and distilled water (300 ml), and the organic layer separated. The aqueous layer was washed twice further with dichloromethane (100 ml x 2), and the organic layers recombined and dried over magnesium sulfate. After the *in vacuo* removal of the dichloromethane, the pale yellow crystals were crushed into a fine powder and left to stir overnight under high vacuum at 40 °C to remove the hexaethyldisiloxane by-product. Recrystallising twice from boiling ethanol (2 x 150 ml) and drying in a vacuum oven overnight at 50 °C afforded the product as a bright white powder.

***<u>BrB6Br - 1-bromo-4-[4-(6-bromohexyl)phenyl]benzene:</u>***
Yield: 20.13 g, 78.5 %.
Melting point: 74.9 – 75.2 °C.
Infrared $\bar{\nu}$ cm-1: 2932, 2856 (sp3 C-H); 1607, 1480 (Ar C=C); 804 (p. disubs. benzene C-H), 647 (C-Br).
$^{1}$H NMR: (300 MHz, Chloroform-d) δ ppm: 1.33 – 1.55 (m, 4 H, -$CH_2$-**<u>CH2</u>**-**<u>CH2</u>**-$CH_2$-$CH_2$-Br), 1.61 - 1.75 (m, 2 H, Ar-$CH_2$-**<u>CH2</u>**-$CH_2$-), 1.89 (quin, J=7.1 Hz, 2 H, -$CH_2$-**<u>CH2</u>**-$CH_2$-Br), 2.62 – 2.77 (m, 2 H, Ar-**<u>CH2</u>**-$CH_2$-), 3.43 (t , J=6.8 Hz, 2 H, -$CH_2$-**<u>CH2</u>**-Br), 7.21 – 7.30 (m, 2 H, Ar), 7.41 – 7.53 (m, 4 H, Ar), 7.53 – 7.61 (m, 2 H, Ar).
$^{13}$C NMR: (75 MHz, Chloroform-d) δ ppm: 28.02, 28.39, 31.18, 32.73, 33.90, 35.43, 121.21, 126.83, 128.56, 128.95, 131.81, 137.42, 140.04, 142.18.

**Synthesis of 4-{[6-(4-(4-bromophenyl)phenyl]hexyl}oxy)-benzaldehyde (BrB6OPhK).**

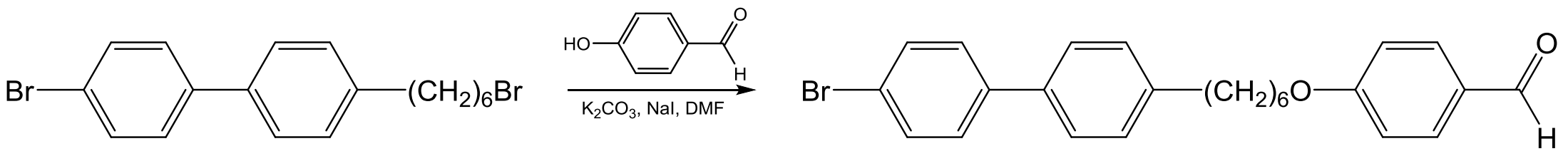

Scheme S3: Synthesis of 4-{[6-(4-(4-bromophenyl)phenyl]hexyl}oxy)benzaldehyde (**BrB6OPhK**).

The synthesis of 4-{[6-(4-(4-bromophenyl)phenyl]hexyl}oxy)benzaldehyde was carried out using Williamson's ether synthesis catalysed *via* an in-situ Finkelstein substitution, as outlined in Scheme 3. Thus, to a stirring solution of 1-bromo-4-[4-(6-bromohexyl)phenyl]benzene (6.083 g, 1.54 x $10^{-2}$ mol, 1.0 mol. eq.) in N,N'-dimethylformamide (30 ml), 4-hydroxybenzaldehyde (1.887 g, 1.55 x $10^{-2}$ mol, 1.0 mol. eq.), potassium carbonate (21.294 g, 1.54 x $10^{-1}$ mol, 10.0 mol. eq.) and sodium iodide (207 mg, 1.38 x $10^{-3}$ mol, 0.1 mol. eq.) was added. The mixture was stirred at 90 °C for 24 h, then left to cool to room temperature before adding to distilled water (350 ml). The resulting cream paste-like precipitate was collected *via* vacuum filtration, added to dichloromethane (50 ml), and washed twice with distilled water (50 ml x 2). The organic layer was dried over magnesium sulfate, and the solvent removed *via* rotary evaporation to yield off-white powder. The powder was twice recrystallised from ethanol (30 ml x 2) and dried *in vacuo* at 50 °C to give the title compound.

***BrB6OPhK** - 4-{[6-(4-(4-bromophenyl)phenyl]hexyl}oxy)benzaldehyde:*
Yield: 5.591 g, 83.1 %.
Melting point: 87.7 – 89.3 °C.
Infrared $\bar{\nu}$ $cm^{-1}$: 2934, 2859 ($sp^3$ C-H); 1691 (C=O); 1600, 1577, 1477 (Ar C=C); 806 (p. disubs. benzene C-H).
$^1H$ NMR: (300 MHz, Chloroform-d) δ ppm: 1.39 – 1.62 (m, 4 H, $-CH_2$-**$CH_2$**-**$CH_2$**-$CH_2$-$CH_2$-O-), 1.71 (quin, J=7.5 Hz, 2 H, Ar-$CH_2$-**$CH_2$**-$CH_2$-), 1.78 - 1.94 (m, 2 H, -$CH_2$-**$CH_2$**-$CH_2$-O-), 2.69 (t, J=7.7 Hz, 2 H, Ar-**$CH_2$**-$CH_2$-), 4.06 (t , J=6.40 Hz, 2 H, -$CH_2$-**$CH_2$**-O-), 7.00 (d, J=8.7 Hz, 2 H, Ar), 7.24 – 7.31 (m, 2 H, Ar), 7.42 – 7.52 (m, 4 H, Ar), 7.53 – 7.62 (m, 2 H, Ar), 7.81 – 7.90 (m, 2 H, Ar), 9.90 (s, 1 H, **HC(O)**-Ar).
$^{13}C$ NMR: (75 MHz, Chloroform-d) δ ppm: 25.86, 28.93, 28.98, 31.27, 35.46, 68.32, 114.75, 121.23, 126.81, 128.54, 128.97, 129.81, 131.82, 131.99, 137.41, 140.01, 142.22, 164.22, 190.78.

**Synthesis of 4-{4-[6-(4-formylphenoxy)hexyl]phenyl}benzonitrile (CB6OPhK).**

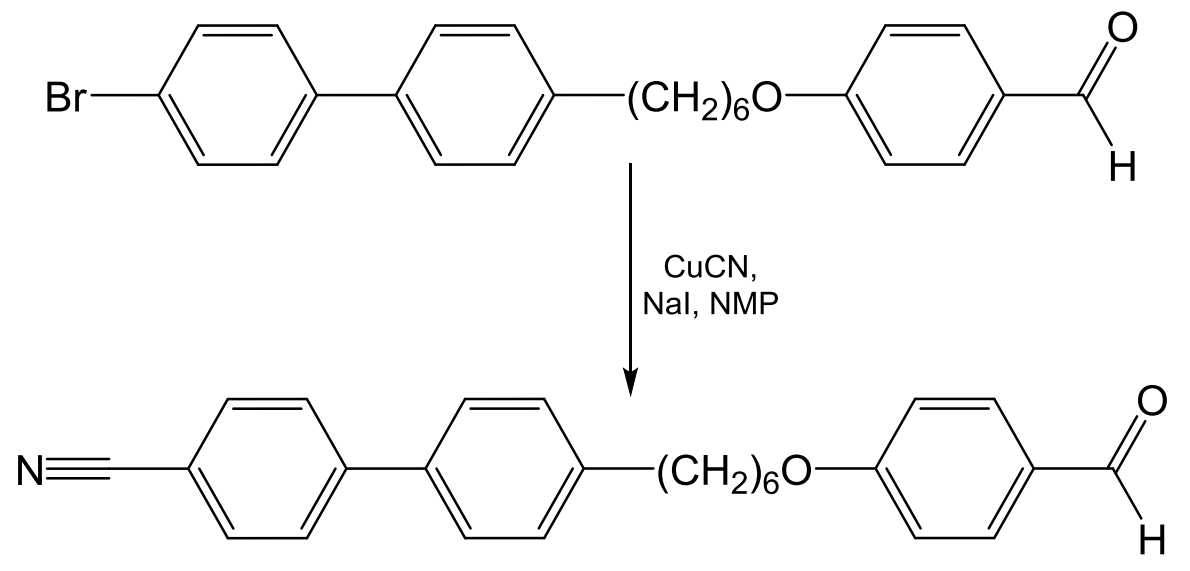

Scheme S4: Synthesis of 4-{4-[6-(4-formylphenoxy)hexyl]phenyl}benzonitrile (**CB6OPhK**).

A Rosenmund-von Braun reaction was used in the cyanation of 4-{[6-(4-(4-bromophenyl)phenyl]hexyl}oxy)benzaldehyde, as shown in Scheme S4 [4]. Copper (I) cyanide (2.230 g, 2.49 x $10^{-2}$ mol, 2.0 mol. eq.), 4-{[6-(4-(4-bromophenyl)phenyl]hexyl}oxy)benzaldehyde (5.407 g, 1.24 x $10^{-2}$ mol, 1.0 mol. eq.), and sodium iodide (1.829 g, 1.22 x $10^{-2}$ mol, 1.0 mol. eq.) were refluxed in N-methyl-2-pyrollidone (50 ml) at 200 °C for 5 h with rapid stirring. The black solution was then cooled to room temperature, and 9 % ammonium hydroxide solution (20 ml) was added, and the mixture stirred for 10 min while bubbling any product gases through a 33% solution of sodium hypochlorite. The dichloromethane layer was collected and washed with brine (100 ml x 2), then with 2M hydrochloric acid (100 ml x 2), and finally with distilled water (100 ml x 2). The organic layer was collected and dried over magnesium sulfate, and the solvent removed using rotary evaporation to give a dark brown liquid. The crude product was purified *via* silica gel chromatography using dichloromethane as an eluent, then twice recrystallised from ethanol (30 ml x 2) and dried *in vacuo* at 50 °C overnight.

***CB6OPhK** - 4-{4-[6-(4-formylphenoxy)hexyl]phenyl}benzonitrile:*
Yield: 3.226 g, 67.8 %.
Melting point: 80.6 – 81.8 °C.
Infrared $\bar{\nu}$ $cm^{-1}$: 2928, 2858 ($sp^3$ C-H); 2223 (C≡N); 1689 (C=O); 1599, 1574, 1493 (Ar C=C); 812 (p. disubs. benzene C-H).
$^1$H NMR: (300 MHz, Chloroform-d) δ ppm: 1.39 – 1.62 (m, 4 H, -$CH_2$-**$CH_2$**-**$CH_2$**-$CH_2$-$CH_2$-O-), 1.72 (quin, J=7.5 Hz, 2 H, Ar-$CH_2$-**$CH_2$**-$CH_2$-), 1.79 - 1.93 (m, 2 H, -$CH_2$-**$CH_2$**-$CH_2$-O-), 2.71 (t, J=7.7 Hz, 2 H, Ar-**$CH_2$**-$CH_2$-), 4.06 (t , J=6.4 Hz, 2 H, -$CH_2$-**$CH_2$**-O-), 6.96 – 7.07 (m, 2 H, Ar), 7.31 (d, J=8.3 Hz, 2 H, Ar), 7.49 – 7.58 (m, 2 H, Ar), 7.66 – 7.78 (m, 4 H, Ar), 7.81 – 7.91 (m, 2 H, Ar), 9.90 (s, 1 H, **HC(O)**-Ar).
$^{13}$C NMR: (75 MHz, Chloroform-d) δ ppm: 25.86, 28.94, 28.98, 31.27, 35.51, 68.28, 110.58, 114.73, 119.06, 127.13, 127.48, 129.19, 129.77, 132.01, 132.59, 136.58, 143.45, 145.54, 164.19, 190.84.

**Synthesis of 4-nitrophenyl-4-(alkoxy)benzoate esters (NxBeO*n*).**

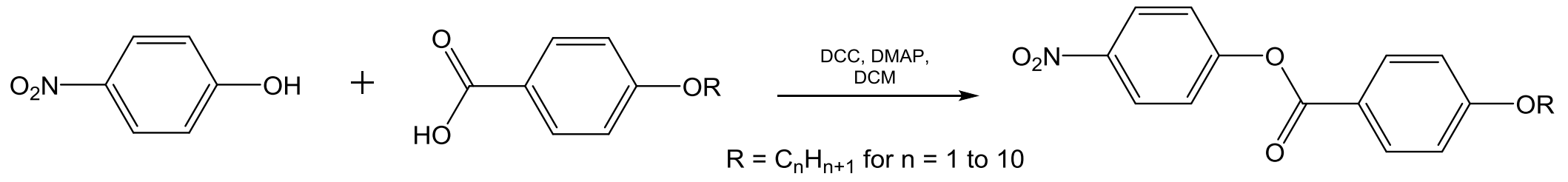

Scheme S5: Synthesis of 4-nitrophenyl-4-(alkoxy)benzoate esters.

4-Nitrophenyl-4-(alkyloxy)benzoate esters were synthesised according to Scheme S5 using a general Steglich ester synthesis as previously reported by Reddy *et al* [5]. As such, the required 4-alkyloxybenzoic acid (1.0 mol. eq.), 4-nitrophenol (1.0 mol. eq.), and dry dichloromethane (150 ml) were added to a conical flask and left to stir for 5 min. 4-Dimethylaminopyridine (0.10 mol. eq.) was added to give an instant yellow solution that was left to stir for 10 min, after which a solution of N,N'-dicyclohexylcarbodiimide (1.3 mol. eq.) in dry dichloromethane (50 ml) was added. Actual quantities used in the synthesis of all ten 4-nitrophenyl-4-(alkyloxy)benzoate intermediates are given in Table S1. The mixtures were left to stir for 12 h at room temperature, after which a white precipitate of N,N'-dicylohexyl urea formed. Most of the precipitate was initially removed *via* vacuum filtration, washing with copious dry dichloromethane, and the remaining through gravity filtration. The clear yellow filtrate was washed three times with a 5% potassium hydroxide solution (3 x 100 ml) and once with distilled water (150 ml), before the organic layer was collected and dried over magnesium sulfate. Dichloromethane was removed *via* rotary evaporation to yield a pale yellow crude solid that was recrystallised from boiling ethanol (150 ml) to give white crystals which were washed with copious ice-cold ethanol, and dried *in vacuo* for 12 h at 50 °C to give the reported yields.

Table S1. Quantities of reagents used in the synthesis of 4-nitrophenyl-4-(alkoxy) benzoate esters

| | **4-Alkoxybenzoic acid used:** | **Quantity of 4-Alkoxybenzoic acid used:** | **Quantity of 4-Nitrophenol used:** | **Quantity of DMAP used:** | **Quantity of DCC used:** |
|---|---|---|---|---|---|
| **n = 1** | 4-Methoxybenzoic acid | 5.009 g, 3.29 x $10^{-2}$ mol, 1.0 eq. | 4.574 g, 3.29 x $10^{-2}$ mol, 1.0 eq. | 437 mg, 3.58 x $10^{-3}$ mol, 0.11 eq. | 8.872 g, 4.27 x $10^{-2}$ mol, 1.3 eq. |
| **n = 2** | 4-Ethoxybenzoic acid | 5.467 g, 3.29 x $10^{-2}$ mol, 1.0 eq. | 4.690 g, 3.37 x $10^{-2}$ mol, 1.0 eq. | 525 mg, 4.30 x $10^{-3}$ mol, 0.13 eq. | 9.356 g, 4.53 x $10^{-2}$ mol, 1.4 eq. |
| **n = 3** | 4-Propoxybenzoic acid | 3.972 g, 2.20 x $10^{-2}$ mol, 1.0 eq. | 3.333 g, 2.40 x $10^{-2}$ mol, 1.1 eq. | 317 mg, 2.59 x $10^{-3}$ mol, 0.12 eq. | 5.846 g, 2.84 x $10^{-2}$ mol, 1.3 eq. |
| **n = 4** | 4-Butoxybenzoic acid | 6.394 g, 3.29 x $10^{-2}$ mol, 1.0 eq. | 4.634 g, 3.33 x $10^{-2}$ mol, 1.0 eq. | 427 mg, 3.50 x $10^{-3}$ mol, 0.11 eq. | 8.740 g, 4.24 x $10^{-2}$ mol, 1.3 eq. |
| **n = 5** | 4-Pentyloxybenzoic acid | 6.911 g, 3.32 x $10^{-2}$ mol, 1.0 eq. | 4.710 g, 3.39 x $10^{-2}$ mol, 1.0 eq. | 411 mg, 3.36 x $10^{-3}$ mol, 0.10 eq. | 8.502 g, 4.12 x $10^{-2}$ mol, 1.2 eq. |
| **n = 6** | 4-Hexyloxybenzoic acid | 1.632 g, 7.34 x $10^{-3}$ mol, 1.0 eq. | 1.140 g, 8.19 x $10^{-3}$ mol, 1.1 eq. | 107 mg, 8.76 x $10^{-4}$ mol, 0.12 eq. | 1.927 g, 9.34 x $10^{-3}$ mol, 1.3 eq. |
| **n = 7** | 4-Heptyloxybenzoic acid | 7.773 g, 3.29 x $10^{-2}$ mol, 1.0 eq. | 4.586 g, 3.30 x $10^{-2}$ mol, 1.0 eq. | 491 mg, 4.02 x $10^{-3}$ mol, 0.12 eq. | 8.503 g, 4.12 x $10^{-3}$ mol, 1.3 eq. |
| **n = 8** | 4-Octyloxybenzoic acid | 4.993 g, 1.99 x $10^{-2}$ mol, 1.0 eq. | 2.851 g, 2.04 x $10^{-2}$ mol, 1.0 eq. | 316 mg, 2.59 x $10^{-3}$ mol, 0.13 eq. | 5.420 g, 2.63 x $10^{-2}$ mol, 1.3 eq. |
| **n = 9** | 4-Nonyloxybenzoic acid | 4.270 g, 1.61 x $10^{-2}$ mol, 1.0 eq. | 2.328 g, 1.67 x $10^{-2}$ mol, 1.0 eq. | 224 mg, 1.83 x $10^{-3}$ mol, 0.11 eq. | 4.291 g, 2.08 x $10^{-2}$ mol, 1.3 eq. |
| **n = 10** | 4-Decyloxybenzoic acid | 4.862 g, 1.75 x $10^{-2}$ mol, 1.0 eq. | 2.741 g, 1.97 x $10^{-2}$ mol, 1.1 eq. | 287 mg, 2.35 x $10^{-3}$ mol, 0.13 eq. | 4.565 g, 2.21 x $10^{-2}$ mol, 1.3 eq. |

***NxBeO1** – 4-Nitrophenyl-4-methoxybenzoate:*

Yield: 7.85 g, 87.3 %.

Melting point: 162.0 – 162.8 °C.

Infrared $\bar{\nu}$ $cm^{-1}$: 3081 ($sp^2$ C-H); 2988, 2944, 2847 ($sp^3$ C-H); 1730 (C=O); 1609, 1583, 1490 (Ar C=C); 1513 ($NO_2$); 849, 832 (p. disubs. benzene C-H). $^1$H NMR: (300 MHz, Chloroform-d) δ ppm: 3.94 (s, 3 H, -Ar-O-**$CH_3$**), 7.03 (d, J=9.0 Hz, 2H, Ar), 7.42 (d, J=9.2 Hz, 2 H, Ar), 8.17 (d, J=9.0 Hz, 2 H, Ar), 8.33 (d, J=9.2 Hz, 2 H, Ar).

$^{13}$C NMR: (75 MHz, Chloroform-d) δ ppm: 55.61, 114.09, 120.72, 122.68, 125.23, 132.54, 145.26, 155.94, 163.92, 164.42.

***NxBeO2** – 4-Nitrophenyl-4-ethoxybenzoate:*

Yield: 8.64 g, 91.4 %.

Melting point: 124.6 – 125.7 °C.

Infrared $\bar{\nu}$ $cm^{-1}$: 3081 ($sp^2$ C-H); 2990 ($sp^3$ C-H); 1727 (C=O); 1610, 1581, 1491 (Ar C=C); 1515 ($NO_2$); 843, 831 (p. disubs. benzene C-H).

$^1$H NMR: (300 MHz, Chloroform-d) δ ppm: 1.49 (t, *J*=7.0 Hz, 3 H, -O-$CH_2$-**$CH_3$**), 4.16 (q, *J*=7.1 Hz, 2 H, -O-**$CH_2$**-$CH_3$), 7.01 (d, *J*=9.0 Hz, 2 H, Ar), 7.42 (d, *J*=9.0 Hz, 2 H, Ar), 8.15 (d, *J*=9.0 Hz, 2 H, Ar), 8.33 (d, *J*=9.0 Hz, 2 H, Ar).

$^{13}$C NMR: (75 MHz, Chloroform-d) δ ppm: 14.66, 63.95, 114.49, 120.45, 122.69, 125.23, 132.54 145.23, 155.97, 163.85, 163.96.

***NxBeO3** – 4-Nitrophenyl-4-propoxybenzoate:*

Yield: 5.33 g, 80.4 %.

Melting point: 63.2 – 63.8 °C.

Infrared $\bar{\nu}$ $cm^{-1}$: 3075 ($sp^2$ C-H); 2973 ($sp^3$ C-H); 1723 (C=O); 1607, 1592, 1493 (Ar C=C); 1514 ($NO_2$); 847 (p. disubs. benzene C-H).

$^1$H NMR: (300 MHz, Chloroform-d) δ ppm: 1.08 (t, *J*=7.4 Hz, 3 H, -$CH_2$-**$CH_3$**), 1.86 (qt, J=7.4, 6.50 Hz, 2 H, -O-$CH_2$-**$CH_2$**-$CH_3$) 4.02 (t, *J*=6.5 Hz, 2 H, -O-**$CH_2$**-$CH_2$-), 6.99 (d, *J*=9.0 Hz, 2 H, Ar), 7.40 (d, *J*=9.4 Hz, 2 H, Ar), 8.13 (d, *J*=9.0 Hz, 2 H, Ar), 8.29 (d, *J*=8.9 Hz, 2 H, Ar).

$^{13}$C NMR: (75 MHz, Chloroform-d) δ ppm: 10.42, 22.42, 69.90, 114.54, 120.44, 122.63, 125.15, 132.49, 145.24, 156.00, 163.91, 164.08.

***NxBeO4** – 4-Nitrophenyl-4-butoxybenzoate:*

Yield: 8.34 g, 80.4 %.

Melting point: 61.2 – 62.6 °C.

Infrared $\bar{\nu}$ cm$^{-1}$: 3086 (sp$^2$ C-H); 2969, 2944, 2877 (sp$^3$ C-H); 1734 (C=O); 1606, 1593, 1490 (Ar C=C); 1527 ($NO_2$); 842, 827 (p. disubs. benzene C-H).

$^1$H NMR: (300 MHz, Chloroform-d) δ ppm: 1.02 (t, *J*=7.4 Hz, 3 H, -$CH_2$-**$CH_3$**), 1.54 (qt, *J*=7.5, 7.2 Hz, 2 H, -$CH_2$-**$CH_2$**-$CH_3$), 1.84 (tt, *J*=7.5, 6.5 Hz, 2 H, -O-$CH_2$-**$CH_2$**-$CH_2$-), 4.08 (t, *J*=6.5 Hz, 2 H, -O-**$CH_2$**-$CH_2$-), 7.01 (d, *J*=9.0 Hz, 2 H, Ar), 7.42 (d, *J*=9.2 Hz, 2 H, Ar), 8.15 (d, *J*=9.0 Hz, 2 H, Ar), 8.33 (d, *J*=9.2 Hz, 2 H, Ar).

$^{13}$C NMR: (75 MHz, Chloroform-d) δ ppm: 13.84, 19.20, 31.10, 68.12, 114.51, 120.37, 122.70, 125.23, 132.53, 145.22, 155.97, 163.98, 164.07.

***NxBeO5** – 4-Nitrophenyl-4-pentoxybenzoate:*

Yield: 7.65 g, 70.0 %.

Melting point: 61.9 – 62.4 °C.

Infrared $\bar{\nu}$ cm$^{-1}$: 3085 (sp$^2$ C-H); 2963, 2944, 2875, 2860 (sp$^3$ C-H); 1733 (C=O); 1607, 1588, 1489 (Ar C=C); 1513 ($NO_2$); 846 (p. disubs. benzene C-H).

$^1$H NMR: (300 MHz, Chloroform-d) δ ppm: 0.97 (t, *J*=7.2 Hz, 3 H, -$CH_2$-**$CH_3$**), 1.33 – 1.56 (m, 4 H, -$CH_2$-**$CH_2$**-**$CH_2$**-$CH_3$), 1.85 (tt, *J*=7.3, 6.6 Hz, 2 H, -O-$CH_2$-**$CH_2$**-$CH_2$-), 4.07 (t, *J*=6.6 Hz, 2 H, -O-**$CH_2$**-$CH_2$-), 7.00 (d, *J*=9.0 Hz, 2 H, Ar), 7.41 (d, J=9.2 Hz, 2 H, Ar), 8.14 (d, *J*=9.0 Hz, 2 H, Ar), 8.31 (d, J=9.2 Hz, 2 H).

$^{13}$C NMR: (75 MHz, Chloroform-d) δ ppm: 14.02, 22.45 28.13, 28.77, 68.43, 114.51, 120.38, 122.67, 125.19, 132.51, 145.21, 155.98, 163.93, 164.07.

***NxBeO6** – 4-Nitrophenyl-4-hexyloxybenzoate:*

Yield: 2.02 g, 80.1 %.

Melting point: 64.7 – 65.5 °C.

Infrared $\bar{\nu}$ cm$^{-1}$: 3075 (sp$^2$ C-H); 2930, 2866 (sp$^3$ C-H); 1738 (C=O); 1606, 1591, 1489 (Ar C=C); 1513 ($NO_2$); 850 (p. disubs. benzene C-H).

$^1$H NMR: (300 MHz, Chloroform-d) δ ppm: 0.93 (t, *J*=7.0 Hz, 3 H, -$CH_2$-**$CH_3$**), 1.28 – 1.58 (m, 6 H, -$CH_2$-**$CH_2$**-**$CH_2$**-**$CH_2$**-$CH_3$), 1.84 (tt, *J*=7.4, 6.6 Hz, 2 H, -O-$CH_2$-**$CH_2$**-$CH_2$-), 4.06 (t, *J*=6.6 Hz, 2 H, -O-**$CH_2$**-$CH_2$-), 7.00 (d, *J*=8.9 Hz, 2 H, Ar), 7.41 (d, J=9.2 Hz, 2 H, Ar), 8.14 (d, *J*=8.9 Hz, 2 H, Ar), 8.31 (d, J=9.2 Hz, 2 H).

$^{13}$C NMR: (75 MHz, Chloroform-d) δ ppm: 14.07, 22.61, 25.67, 29.04, 31.56, 68.44, 114.51, 120.35, 122.69, 125.20, 132.51, 145.18, 155.98, 163.95, 164.07.

***NxBeO7** – 4-Nitrophenyl-4-heptyloxybenzoate:*

Yield: 9.34 g, 79.4 %.

Melting point: 57.9 °C (Cr-N), 60.9 °C (N-I).

Infrared $\bar{\nu}$ cm$^{-1}$: 3080 (sp$^2$ C-H); 2925, 2860 (sp$^3$ C-H); 1732 (C=O); 1608, 1590, 1490 (Ar C=C); 1519 ($NO_2$); 848, 837 (p. disubs. benzene C-H).

$^1$H NMR: (300 MHz, Chloroform-d) δ ppm: 0.92 (t, *J*=6.8 Hz, 3 H, -$CH_2$-**$CH_3$**), 1.25 – 1.57 (m, 8 H, -$CH_2$-**$CH_2$**-**$CH_2$**-**$CH_2$**-**$CH_2$**-$CH_3$), 1.85 (tt, *J*=7.3, 6.6 Hz, 2 H, -O-$CH_2$-**$CH_2$**-$CH_2$-), 4.07 (t, *J*=6.5 Hz, 2 H, -O-**$CH_2$**-$CH_2$-), 7.01 (d, *J*=8.9 Hz, 2 H, Ar), 7.42 (d, *J*=9.2 Hz, 2 H, Ar), 8.15 (d, *J*=9.0 Hz, 2 H, Ar), 8.33 (d, *J*=9.2 Hz, 2 H, Ar).

$^{13}C$ NMR: (75 MHz, Chloroform-d) δ ppm: 14.12, 22.63, 25.95, 29.05, 29.08, 31.78, 68.45, 114.51, 120.35, 122.70, 125.23, 132.53, 145.21, 155.98, 163.98, 164.07.

***NxBeO8** – 4-Nitrophenyl-4-octyloxybenzoate*:
Yield: 5.43 g, 73.5 %.
Melting point: 50.5 °C (Cr-SmA), 60.6 °C (SmA-N), 67.3 °C (N-I).
Infrared $\bar{\nu}$ $cm^{-1}$: 3085 ($sp^2$ C-H); 2927, 2871 ($sp^3$ C-H); 1736 (C=O); 1605, 1591, 1488 (Ar C=C); 1510 ($NO_2$); 843 (p. disubs. benzene C-H).
$^{1}H$ NMR: (300 MHz, Chloroform-d) δ ppm: 0.92 (t, *J*=7.2 Hz, 3 H, -$CH_2$-**$CH_3$**), 1.23 – 1.58 (m, 10 H, -$CH_2$-**$CH_2$**-**$CH_2$**-**$CH_2$**-**$CH_2$**-**$CH_2$**-$CH_3$), 1.85 (tt, *J*=7.0, 6.6 Hz, 2 H, -O-$CH_2$-**$CH_2$**-$CH_2$-), 4.08 (t, *J*=6.6 Hz, 2 H, -O-**$CH_2$**-$CH_2$-), 7.01 (d, *J*=9.0 Hz, 2 H, Ar), 7.42 (d, *J*=9.0 Hz, 2 H, Ar), 8.15 (d, *J*=8.7 Hz, 2 H, Ar), 8.33 (d, *J*=9.4 Hz, 2 H, Ar).
$^{13}C$ NMR: (75 MHz, Chloroform-d) δ ppm: 14.05, 22.63, 25.97, 29.06, 29.19, 29.30, 31.78, 68.46, 114.54, 120.45, 122.63, 125.18, 132.49, 145.28, 155.99, 163.93, 164.09.

***NxBeO9** – 4-Nitrophenyl-4-nonyloxybenzoate*:
Yield: 5.10 g, 82.2 %.
Melting point: 53.8 °C (Cr-SmA), 67.9 °C (SmA-I).
Infrared $\bar{\nu}$ $cm^{-1}$: 3087 ($sp^2$ C-H); 2964, 2953, 2943, 2919, 2853 ($sp^3$ C-H); 1733 (C=O); 1607, 1594, 1490 (Ar C=C); 1511 ($NO_2$); 842, 831 (p. disubs. benzene C-H).
$^{1}H$ NMR: (300 MHz, Chloroform-d) δ ppm: 0.92 (t, *J*=7.2 Hz, 3 H, -$CH_2$-**$CH_3$**), 1.31 – 1.58 (m, 12 H, -$CH_2$-**$CH_2$**-**$CH_2$**-**$CH_2$**-**$CH_2$**-**$CH_2$**-**$CH_2$**-$CH_3$), 1.86 (tt, *J*=7.2, 6.5 Hz, 2 H, -O-$CH_2$-**$CH_2$**-$CH_2$-), 4.08 (t, *J*=6.5 Hz, 2 H, -O-**$CH_2$**-$CH_2$-), 7.02 (d, *J*=8.9 Hz, 2 H, Ar), 7.43 (d, *J*=8.7 Hz, 2 H, Ar), 8.15 (d, *J*=9.0 Hz, 2 H, Ar), 8.31 (d, *J*=9.2 Hz, 2 H, Ar).
$^{13}C$ NMR: (75 MHz, Chloroform-d) δ ppm: 13.99, 22.60, 25.91, 29.02, 29.18, 29.30, 29.45, 31.81, 68.43, 114.51, 120.35, 122.59, 125.09, 132.44, 145.18, 155.96, 163.93, 164.06.

***NxBeO10** – 4-Nitrophenyl 4-decyloxybenzoate*:
Yield: 5.94 g, 85.0 %.
Melting point: 58.9 °C (Cr-SmA), 76.4 (SmA-I) °C.
Infrared $\bar{\nu}$ $cm^{-1}$: 3086 ($sp^2$ C-H); 2954, 2942, 2918, 2873, 2852 ($sp^3$ C-H); 1732 (C=O); 1607, 1594, 1491 (Ar C=C); 1511 ($NO_2$); 841 (p. disubs. benzene C-H).
$^{1}H$ NMR: (300 MHz, Chloroform-d) δ ppm: 0.90 (t, *J*=6.8 Hz, 3 H, -$CH_2$-**$CH_3$**), 1.20 – 1.57 (m, 14 H, -$CH_2$-**$CH_2$**-**$CH_2$**-**$CH_2$**-**$CH_2$**-**$CH_2$**-**$CH_2$**-**$CH_2$**-$CH_3$), 1.85 (tt, *J*=7.5, 6.5 Hz, 2 H, -O-$CH_2$-**$CH_2$**-$CH_2$-), 4.07 (t, *J*=6.5 Hz, 2 H, -O-**$CH_2$**-$CH_2$-), 7.00 (d, *J*=8.9 Hz, 2 H, Ar), 7.42 (d, *J*=9.2 Hz, 2 H, Ar), 8.15 (d, *J*=8.9 Hz, 2 H, Ar), 8.33 (d, *J*=9.0 Hz, 2 H, Ar).
$^{13}C$ NMR: (75 MHz, Chloroform-d) δ ppm: 14.15, 22.71, 25.99, 29.08, 29.35, 29.38, 29.57, 31.91, 68.45, 114.51, 120.36, 122.68, 125.21, 132.51, 145.21, 155.98, 163.95, 164.07.

**Synthesis of 4-aminophenyl-4’-(alkoxy)benzoate esters (H2NBeOn)**

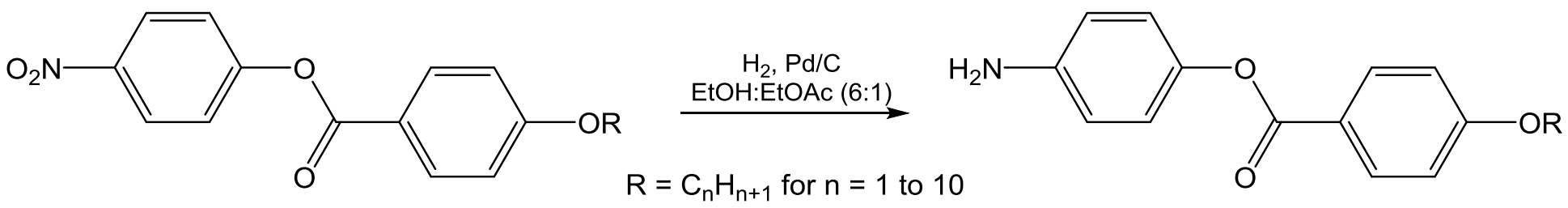

Scheme S6: Synthesis of 4-aminophenyl-4’-(alkoxy)benzoate esters.

Selective reduction of 4-nitrophenyl-4’-(alkoxy)benzoate esters over a palladium/activated charcoal catalyst gave the 4-aminophenyl-4’-(alkoxy)benzoate intermediates, as outlined in Scheme S6. In a flask, the required 4-nitrophenyl-4’-(alkoxy)benzoate (1 mol. eq.) was dissolved in ethyl acetate (50 ml) and ethanol (300 ml), with 10 % Pd/C, 53% wet (0.01 mol. eq. of Pd), and flushed with argon before evacuating under vacuum. The actual quantities used for each reduction are given in Table S2. Following de-gassing, hydrogen gas was bubbled into the vessel *via* a balloon at ambient pressure and temperature. After 12 h, solid Pd/C was filtered out through a Celite plug and washed with ethyl acetate (300 ml), and the solvent removed *via* rotary evaporation to give the desired 4-aminophenyl-4-(alkoxy)benzoate ester. The reported yields of the beige crystals are given below after recrystallisation from boiling ethanol (30 ml/g of ester), and drying overnight *in vacuo* at 50 °C.

Table S2: Quantities of reagents used in the synthesis of 4-amino phenyl-4-(alkyloxy) benzoate intermediates.

| | **4-Nitrophenyl 4’-(alkyloxy)benzoate used:** | **Quantity of 4-nitrophenyl 4’-(alkyloxy)benzoate used:** | **Quantity of 10% Pd/C, 53 % wet used:** |
|---|---|---|---|
| n = 1 | 4-Nitrophenyl 4’-methoxybenzoate | 7.018 g, 2.57 x $10^{-2}$ mol, 1.0 eq. | 705 mg, 3.11 x $10^{-4}$ mol, 0.012 eq. |
| n = 2 | 4-Nitrophenyl 4’-ethoxybenzoate | 8.014 g, 2.79 x $10^{-2}$ mol, 1.0 eq. | 774 mg, 3.42 x $10^{-4}$ mol, 0.012 eq. |
| n = 3 | 4-Nitrophenyl 4’-propoxybenzoate | 4.626 g, 1.54 x $10^{-2}$ mol, 1.0 eq. | 441 mg, 1.95 x $10^{-4}$ mol, 0.013 eq. |
| n = 4 | 4-Nitrophenyl 4’-butoxybenzoate | 7.853 g, 2.49 x $10^{-2}$ mol, 1.0 eq. | 701 mg, 3.10 x $10^{-4}$ mol, 0.012 eq. |
| n = 5 | 4-Nitrophenyl 4’-pentoxybenzoate | 7.012 g, 2.13 x $10^{-2}$ mol, 1.0 eq. | 584 mg, 2.58 x $10^{-4}$ mol, 0.012 eq. |
| n = 6 | 4-Nitrophenyl 4’-hexyloxybenzoate | 1.722 g, 5.01 x $10^{-3}$ mol, 1.0 eq. | 142 mg, 6.27 x $10^{-5}$ mol, 0.013 eq. |
| n = 7 | 4-Nitrophenyl 4’-heptyloxybenzoate | 8.721 g, 2.44 x $10^{-2}$ mol, 1.0 eq. | 681 mg, 3.11 x $10^{-4}$ mol, 0.012 eq. |
| n = 8 | 4-Nitrophenyl 4’-octyloxybenzoate | 4.978 g, 1.34 x $10^{-2}$ mol, 1.0 eq. | 378 mg, 1.67 x $10^{-4}$ mol, 0.012 eq. |
| n = 9 | 4-Nitrophenyl 4’-nonyloxybenzoate | 4.632 g, 1.20 x $10^{-2}$ mol, 1.0 eq. | 338 mg, 1.49 x $10^{-4}$ mol, 0.012 eq. |
| n = 10 | 4-Nitrophenyl 4’-decyloxybenzoate | 5.573 g, 1.39 x $10^{-2}$ mol, 1.0 eq. | 392 mg, 1.73 x $10^{-4}$ mol, 0.012 eq. |

***H$_2$NBeO1** – 4-Aminophenyl-4'-methoxybenzoate:*

Yield: 6.14 g, 98.3 %.

Melting point: 140.9 °C.

Infrared $\bar{\nu}$ cm$^{-1}$: 3464, 3379 (N-H); 3019 (sp$^2$ C-H); 2977, 2922, 2843 (sp$^3$ C-H); 1713 (C=O); 1606, 1578, 1509 (Ar C=C); 1264 (Ar C-O); 843, 821 (p. disubs. benzene C-H).

$^1$H NMR: (300 MHz, Chloroform-d) δ ppm: 3.70 (br. s., 2 H, **$H_2N$**-Ar-), 3.90 (s, 3H, -O-**$CH_3$**), 6.71 (d, *J*=8.7 Hz, 2 H, Ar), 6.91 - 7.11 (m, 4 H, Ar), 8.17 (d, *J*=8.8 Hz, 2 H, Ar).

$^{13}$C NMR: (75 MHz, Chloroform-d) δ ppm: 55.52, 113.81, 115.68, 122.16, 122.35, 132.20, 143.10, 144.31, 163.77, 165.51.

***H$_2$NBeO2** – 4-Aminophenyl-4'-ethoxybenzoate:*

Yield: 7.01 g, 97.7 %.

Melting point: 152.2 °C.

Infrared $\bar{\nu}$ cm$^{-1}$: 3493, 3395 (N-H); 2978, 2933 (sp$^3$ C-H); 1713 (C=O); 1605, 1511 (Ar C=C); 1257 (Ar C-O); 848, 819 (p. disubs. benzene C-H).

$^1$H NMR: (300 MHz, Chloroform-d) δ ppm: 1.48 (t, J=7.0 Hz, 3 H, -O-$CH_2$-**$CH_3$**), 3.80 (br. s., 2 H, **$H_2N$**-Ar-), 4.13 (q, J=7.0 Hz, 2 H, -O-**$CH_2$**-$CH_3$), 6.72 (d, *J*=8.7 Hz, 2 H, Ar), 6.92 - 7.07 (m, 4 H, Ar), 8.15 (d, *J*=8.9 Hz, 2 H, Ar).

$^{13}$C NMR: (75 MHz, Chloroform-d) δ ppm: 14.71, 63.79, 114.19, 115.69, 121.90, 122.38, 132.20, 143.16, 144.17, 163.16, 165.53.

***H$_2$NBeO3** – 4-Aminophenyl-4'-propoxybenzoate:*

Yield: 4.01 g, 96.2 %.

Melting point: 133.4 °C.

Infrared $\bar{\nu}$ cm$^{-1}$: 3476, 3388 (N-H); 2966, 2912, 2877 (sp$^3$ C-H); 1719 (C=O); 1604, 1510, 1472 (Ar C=C); 1262 (Ar C-O); 847, 826 (p. disubs. benzene C-H).

$^1$H NMR: (300 MHz, Chloroform-d) δ ppm: 1.07 (t, J=7.4 Hz, 3 H, -$CH_2$-**$CH_3$**), 1.86 (qt, J=7.3, 6.5 Hz, 2 H, -O-$CH_2$-**$CH_2$**-$CH_3$), 3.58 (br. s., 2 H, **$H_2N$**-Ar-), 3.99 (t, J=6.5 Hz, 2 H, -O-**$CH_2$**-$CH_2$-), 6.69 (d, *J*=8.7 Hz, 2 H, Ar), 6.89 - 7.06 (m, 4 H, Ar), 8.15 (d, *J*=8.9 Hz, 2 H, Ar).

$^{13}$C NMR: (75 MHz, Chloroform-d) δ ppm: 10.50, 22.47, 69.79, 114.27, 115.70, 121.83, 122.34, 132.19, 143.07, 144.36, 163.30, 165.62.

***H$_2$NBeO4** – 4-Aminophenyl-4'-butoxybenzoate:*

Yield: 7.00 g, 98.5 %.

Melting point: 110.0 °C.

Infrared $\bar{\nu}$ cm$^{-1}$: 3453, 3365 (N-H); 2953, 2939, 2875 (sp$^3$ C-H); 1708 (C=O); 1604, 1578, 1507, 1476 (Ar C=C); 1252 (Ar C-O); 852, 826 (p. disubs. benzene C-H).

$^1$H NMR: (300 MHz, Chloroform-d) δ ppm: 1.02 (t, *J*=7.4 Hz, 3 H, -$CH_2$-**$CH_3$**), 1.54 (tq, *J*=7.4, 7.3 Hz, 2 H, -$CH_2$-**$CH_2$**-$CH_3$-), 1.83 (tt, *J*=7.4, 6.4 Hz, 2 H, -O-$CH_2$-**$CH_2$**-$CH_2$-), 3.69 (br. s., 2 H, **$H_2N$**-Ar-), 4.06 (t, *J*=6.4 Hz, 2 H, -O-**$CH_2$**-$CH_2$-), 6.71 (d, *J*=8.7 Hz, 2 H, Ar), 6.92 - 7.07 (m, 4 H, Ar), 8.15 (d, *J*=8.7 Hz, 2 H, Ar).

$^{13}$C NMR: (75 MHz, Chloroform-d) δ ppm: 13.87, 19.22, 31.16, 67.98, 114.24, 115.69, 121.85, 122.37, 132.18, 143.16, 144.22, 163.39, 165.56.

***H$_2$NBeO5** – 4-Aminophenyl-4'-pentoxybenzoate*:
Yield: 6.16 g, 96.6 %.
Melting point: 94.0 °C.
Infrared $\bar{\nu}$ cm$^{-1}$: 3466, 3378 (N-H); 2948, 2867 (sp$^3$ C-H); 1718 (C=O); 1603, 1579, 1511 (Ar C=C); 1256 (Ar C-O); 823 (p. disubs. benzene C-H).
$^1$H NMR: (300 MHz, Chloroform-d) δ ppm: 0.97 (t, *J*=7.0 Hz, 3 H, -CH$_2$-**CH$_3$**), 1.34 – 1.57 (m, 4 H, -CH$_2$-**CH$_2$**-**CH$_2$**-CH$_3$), 1.85 (tt, *J*=7.3, 6.5 Hz, 2 H, -O-CH$_2$-**CH$_2$**-CH$_2$-), 3.69 (br. s., 2 H, **H$_2$N**-Ar-), 4.05 (t, *J*=6.5 Hz, 2 H, -O-**CH$_2$**-CH$_2$-), 6.72 (d, *J*=8.7 Hz, 2 H, Ar), 6.91 - 7.07 (m, 4 H, Ar), 8.13 (d, *J*=8.9 Hz, 2 H, Ar).
$^{13}$C NMR: (75 MHz, Chloroform-d) δ ppm: 14.07, 22.47, 28.15, 28.82, 68.28, 114.22, 115.70, 121.82, 122.38, 132.19, 143.16, 144.17, 163.37, 165.56.

***H$_2$NBeO6** – 4-Aminophenyl-4'-hexyloxybenzoate:*
Yield: 1.49 g, 94.5 %.
Melting point: 78.5 °C.
Infrared $\bar{\nu}$ cm$^{-1}$: 3461, 3374 (N-H); 2955, 2929, 2869 (sp$^3$ C-H); 1713 (C=O); 1605, 1581, 1512 (Ar C=C); 1253 (Ar C-O); 843, 826 (p. disubs. benzene C-H).
$^1$H NMR: (300 MHz, Chloroform-d) δ ppm: 0.93 (t, *J*=7.0 Hz, 3 H, -CH$_2$-**CH$_3$**), 1.31 – 1.57 (m, 6 H, -CH$_2$-**CH$_2$**-**CH$_2$**-**CH$_2$**-CH$_3$), 1.83 (tt, *J*=7.2, 6.5 Hz, 2 H, -O-CH$_2$-**CH$_2$**-CH$_2$-), 3.75 (br. s., 2 H, **H$_2$N**-Ar-), 4.04 (t, *J*=6.5 Hz, 2 H, -O-**CH$_2$**- -CH$_2$-), 6.70 (d, *J*=8.9 Hz, 2 H, Ar), 6.91 - 7.04 (m, 4 H, Ar), 8.13 (d, *J*=8.9 Hz, 2 H, Ar).
$^{13}$C NMR: (75 MHz, Chloroform-d) δ ppm: 14.03, 22.59, 25.67, 29.08, 31.55, 68.30, 114.23, 115.68, 121.87, 122.35, 132.17, 143.21, 144.15, 163.38, 165.51.

***H$_2$NBeO7** – 4-Aminophenyl-4'-heptyloxybenzoate*:
Yield: 7.77 g, 97.3 %.
Melting point: 85.7 °C.
Infrared $\bar{\nu}$ cm$^{-1}$: 3463, 3375 (N-H); 2949, 2926, 2868 (sp$^3$ C-H); 1714 (C=O); 1605, 1581, 1511 (Ar C=C); 1254 (Ar C-O); 850, 826 (p. disubs. benzene C-H).
$^1$H NMR: (300 MHz, Chloroform-d) δ ppm: 0.93 (t, *J*=6.6 Hz, 3 H, -CH$_2$-**CH$_3$**), 1.30 – 1.53 (m, 8 H, -CH$_2$-**CH$_2$**-**CH$_2$**-**CH$_2$**-**CH$_2$**-CH$_3$), 1.84 (tt, *J*=7.5, 6.5 Hz, 2 H, -O-CH$_2$-**CH$_2$**-CH$_2$-), 3.76 (br. s, 2 H, **H$_2$N**-Ar-), 4.05 (t, *J*=6.5 Hz, 2 H, -O-**CH$_2$**- -CH$_2$-), 6.72 (d, *J*=8.9 Hz, 2 H, Ar), 6.94 – 7.04 (m, 4 H, Ar), 8.15 (d, *J*=8.9 Hz, 2 H, Ar).
$^{13}$C NMR: (75 MHz, Chloroform-d) δ ppm: 14.15, 22.64, 25.97, 29.07, 29.13, 31.80, 68.30, 114.22, 115.69, 121.82, 122.38, 132.19, 143.15, 144.20, 163.38, 165.56.

***H$_2$NBeO8** – 4-Aminophenyl-4'-octyloxybenzoate*:
Yield: 4.37 g, 95.5 %.
Melting point: 93.3 °C.
Infrared $\bar{\nu}$ cm$^{-1}$: 3468, 3374 (N-H); 2958, 2934, 2916, 2850 (sp$^3$ C-H); 1711 (C=O); 1604, 1577, 1510 (Ar C=C); 1254 (Ar C-O); 844, 823 (p. disubs. benzene C-H).
$^1$H NMR: (300 MHz, Chloroform-d) δ ppm: 0.92 (t, *J*=7.0 Hz, 3 H, -CH$_2$-**CH$_3$**), 1.22 – 1.60 (m, 10 H, -CH$_2$-**CH$_2$**-**CH$_2$**-**CH$_2$**-**CH$_2$**-**CH$_2$**-CH$_3$), 1.84 (tt, *J*=7.0, 6.6 Hz, 2 H, -O-CH$_2$-**CH$_2$**-CH$_2$-), 3.69 (br. s, 2 H, **H$_2$N**-Ar-), 4.05 (t, *J*=6.6 Hz, 2 H, -O-**CH$_2$**- -CH$_2$-), 6.72 (d, *J*=8.9 Hz, 2 H, Ar), 6.92 – 7.06 (m, 4 H, Ar), 8.15 (d, *J*=8.9 Hz, 2 H, Ar).

$^{13}C$ NMR: (75 MHz, Chloroform-d) δ ppm: 14.15, 22.69, 26.02, 29.13, 29.26, 29.36, 31.83, 68.30, 114.22, 115.69, 121.82, 122.38, 132.19, 143.16, 144.19, 163.38, 165.56.

***$H_2$NBeO9** – 4-Aminophenyl-4'-nonyloxybenzoate:*
Yield: 3.99 g, 93.5 %.
Melting point: 90.3 °C.
Infrared $\bar{\nu}$ $cm^{-1}$: 3462, 3374 (N-H); 2947, 2918, 2868, 2849 ($sp^3$ C-H); 1715 (C=O); 1605, 1580, 1511 (Ar C=C); 1257 (Ar C-O); 848, 824 (p. disubs. benzene C-H).
$^{1}H$ NMR: (300 MHz, Chloroform-d) δ ppm: 0.92 (t, *J*=7.2 Hz, 3 H, -$CH_2$-**$CH_3$**), 1.26 – 1.51 (m, 12 H, -$CH_2$-**$CH_2$**-**$CH_2$**-**$CH_2$**-**$CH_2$**-**$CH_2$**-**$CH_2$**-$CH_3$), 1.84 (tt, *J*=7.2, 6.6 Hz, 2 H, -O-$CH_2$-**$CH_2$**-$CH_2$-), 3.74 (br. s, 2 H, **$H_2N$**-Ar-), 4.05 (t, *J*=6.6 Hz, 2 H, -O-**$CH_2$**- -$CH_2$-), 6.72 (d, *J*=8.5 Hz, 2 H, Ar), 6.90 – 7.10 (m, 4 H, Ar), 8.15 (d, *J*=9.0 Hz, 2 H, Ar).
$^{13}C$ NMR: (75 MHz, Chloroform-d) δ ppm: 14.17, 22.71, 26.00, 29.12, 29.29, 29.40, 29.55, 31.90, 68.30, 114.22, 115.70, 121.80, 122.38, 132.19, 143.16, 144.17, 163.37, 165.57.

***$H_2$NBeO10** – 4-Aminophenyl-4'-decyloxybenzoate:*
Yield: 4.78 g, 92.8 %.
Melting point: 91.5 °C.
Infrared $\bar{\nu}$ $cm^{-1}$: 3457, 3371 (N-H); 2953, 2919, 2850 ($sp^3$ C-H); 1712 (C=O); 1605, 1579, 1511 (Ar C=C); 1256 (Ar C-O); 846, 826 (p. disubs. benzene C-H)
$^{1}H$ NMR: (300 MHz, Chloroform-d) δ ppm: 0.93 (t, *J*=7.0 Hz, 3 H, -$CH_2$-**$CH_3$**), 1.27 – 1.62 (m, 14 H, -$CH_2$-**$CH_2$**-**$CH_2$**-**$CH_2$**-**$CH_2$**-**$CH_2$**-**$CH_2$**-**$CH_2$**-$CH_3$), 1.85 (tt, *J*=7.0, 6.5 Hz, 2 H, -O-$CH_2$-**$CH_2$**-$CH_2$-), 3.70 (br. s, 2 H, **$H_2N$**-Ar-), 4.05 (t, *J*=6.5 Hz, 2 H, -O-**$CH_2$**-$CH_2$-), 6.71 (d, *J*=9.0 Hz, 2 H, Ar), 6.93 – 7.07 (m, 4 H, Ar), 8.16 (d, *J*=8.9 Hz, 2 H, Ar).
$^{13}C$ NMR: (75 MHz, Chloroform-d) δ ppm: 14.20, 22.74, 23.01, 26.03, 29.14, 29.38, 29.42, 29.61, 31.95, 68.30, 114.24, 115.68, 121.84, 122.37, 132.19, 143.12, 144.28, 163.39, 165.56.

**Synthesis of 4-[{[4-({6-[4-(4-Cyanophenyl)phenyl]hexyl}oxy)phenyl]-methylidene} amino]phenyl 4-alkoxybenzoate dimers (CB6OIBeO*n*).**

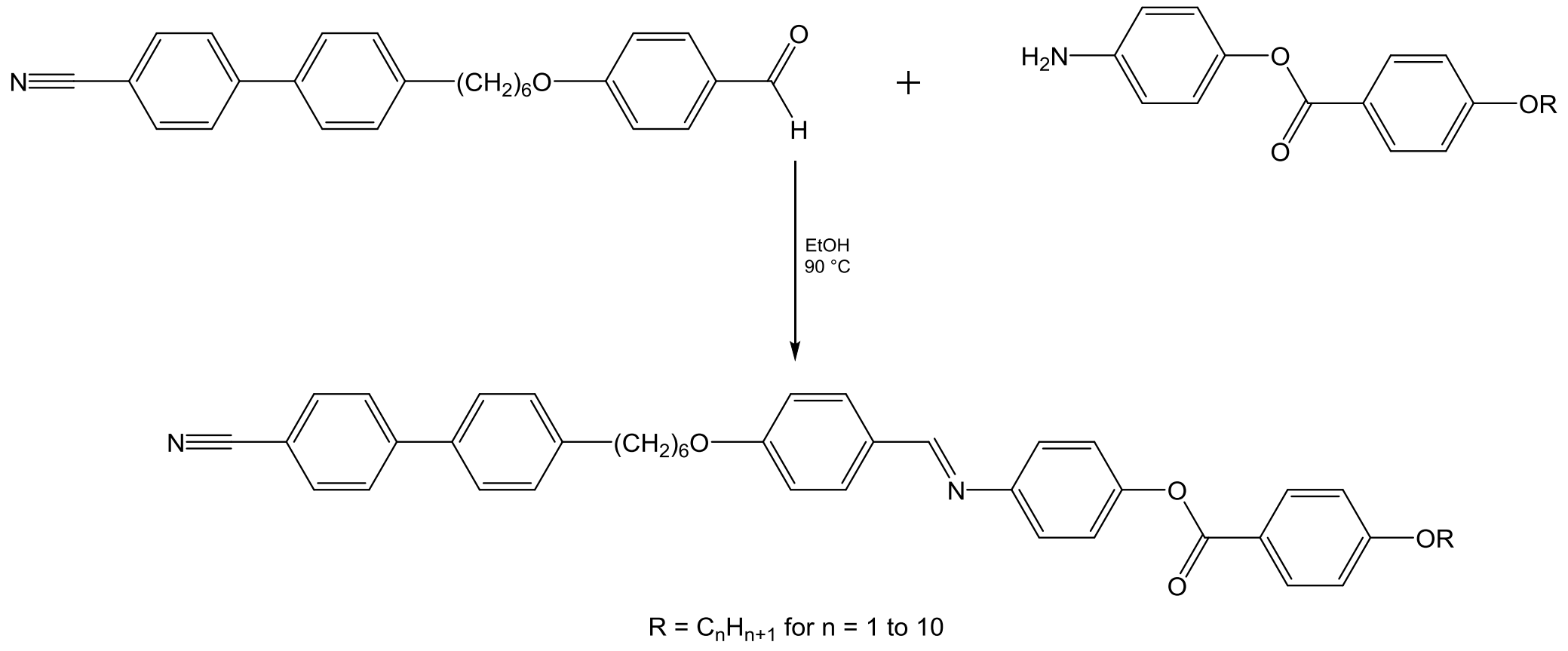

Scheme S7: Synthesis of 4-[{[4-({6-[4-(4-cyanophenyl)phenyl]hexyl}oxy)phenyl] methylidene}amino]phenyl 4-alkoxybenzoate dimers.

The synthesis of 4-[{[4-({6-[4-(4-cyanophenyl)phenyl]hexyl}oxy)phenyl]methylidene} amino]phenyl 4-alkoxybenzoate dimers followed a general Schiff base synthesis, as outlined in Scheme S7. 4-{4-[6-(4-formylphenoxy)hexyl]phenyl}benzonitrile (1 mol. eq.) and the desired 4-aminophenyl-4-(alkoxy)benzoate ester (1 mol. eq.) were combined in ethanol (10 ml) and refluxed with stirring for 2 h. Exact quantities of reagents used are given in Table S3. The pale yellow solutions were left to cool to room temperature, and then to ~0 °C for 30 min. The pale yellow crystals were collected *via* vacuum filtration and recrystallised from boiling ethyl acetate (10 ml) with petroleum ether 40-60 as a co-solvent (5 to 10 drops), with cooling overnight to -20 °C. Recrystallisation was repeated three times, and the compounds dried *in vacuo* overnight at 50 °C to give the dimers in the reported yields. It was not possible to obtain 13C NMR spectra for the complete series because the compounds degraded during the measurement.

Table S3. Quantities of reagents used in the synthesis of 4-[{[4-({6-[4-(4-cyanophenyl) phenyl]hexyl}oxy)phenyl]methylidene}amino]phenyl 4-alkoxybenzoate dimers

| | **4-Aminophenyl 4-(alkyloxy)benzoate used:** | **Quantity of 4-Aminophenyl 4-(alkoxy)benzoate used:** | **Quantity of CB6OIPhK used:** |
|---|---|---|---|
| **n = 1** | 4-Aminophenyl 4-methoxybenzoate | 192 mg, 7.89 x $10^{-4}$ mol, 1.0 eq. | 302 mg, 7.88 x $10^{-4}$ mol, 1.0 eq. |
| **n = 2** | 4-Aminophenyl 4-ethoxybenzoate | 202 mg, 7.85 x $10^{-4}$ mol, 1.0 eq. | 298 mg, 7.77 x $10^{-4}$ mol, 1.0 eq. |
| **n = 3** | 4-Aminophenyl 4-propoxybenzoate | 210 mg, 7.74 x $10^{-4}$ mol, 1.0 eq. | 301 mg, 7.85 x $10^{-4}$ mol, 1.0 eq. |
| **n = 4** | 4-Aminophenyl 4-butoxybenzoate | 226 mg, 7.92 x $10^{-4}$ mol, 1.0 eq. | 300 mg, 7.82 x $10^{-4}$ mol, 1.0 eq. |
| **n = 5** | 4-Aminophenyl 4-pentoxybenzoate | 232 mg, 7.75 x $10^{-4}$ mol, 1.0 eq. | 304 mg, 7.93 x $10^{-4}$ mol, 1.0 eq. |
| **n = 6** | 4-Aminophenyl 4-hexyloxybenzoate | 247 mg, 7.88 x $10^{-4}$ mol, 1.0 eq. | 297 mg, 7.74 x $10^{-4}$ mol, 1.0 eq. |
| **n = 7** | 4-Aminophenyl-4-heptyloxybenzoate | 256 mg, 7.82 x $10^{-4}$ mol, 1.0 eq. | 295 mg, 7.69 x $10^{-4}$ mol, 1.0 eq. |
| **n = 8** | 4-Aminophenyl-4-octyloxybenzoate | 264 mg, 7.73 x $10^{-4}$ mol, 1.0 eq. | 301 mg, 7.85 x $10^{-4}$ mol, 1.0 eq. |
| **n = 9** | 4-Aminophenyl-4-nonyloxybenzoate | 279 mg, 7.85 x $10^{-4}$ mol, 1.0 eq. | 303 mg, 7.90 x $10^{-4}$ mol, 1.0 eq. |
| **n = 10** | 4-Aminophenyl-4-decyloxybenzoate | 285 mg, 7.71 x $10^{-4}$ mol, 1.0 eq. | 302 mg, 7.88 x $10^{-4}$ mol, 1.0 eq. |

***CB6OIBeO1*** *- 4-[{[4-({6-[4-(4-Cyanophenyl)phenyl]hexyl}oxy)phenyl]methylidene}amino] phenyl 4-methoxybenzoate:*

Yield: 287 mg, 59.9 %.

Infrared $\bar{\nu}$ $cm^{-1}$: 3022 ($sp^2$ C-H); 2929, 2853 ($sp^3$ C-H); 2221 (C≡N); 1723 (C=O); 1603, 1572, 1511, 1495 (Ar C=C); 1252 (Ar C-O); 841, 828 (p. disubs. benzene C-H).

$^1$H NMR (300 MHz, DMSO-$d_6$) δ ppm: 1.30 - 1.54 (m, 4 H, -$CH_2$-$CH_2$-**$CH_2$**-**$CH_2$**-$CH_2$-$CH_2$-O-), 1.63 (quin, *J*=7.5, 2 H, -$CH_2$-**$CH_2$**-$CH_2$-$CH_2$-$CH_2$-$CH_2$-O-), 1.74 (tt, *J*=7.2, 6.4 Hz, 2 H, -$CH_2$-$CH_2$-$CH_2$-$CH_2$-**$CH_2$**-$CH_2$-O-), 2.64 (t, *J*=7.5 Hz, 2 H, -Ar-**$CH_2$**-$CH_2$-), 3.88 (s, 3 H, -Ar-O-**$CH_3$**), 4.04 (t, *J*=6.4 Hz, 2 H, -$CH_2$-**$CH_2$**-O-), 7.05 (d, *J*=8.7 Hz, 2 H, Ar), 7.14 (d, *J*=9.0 Hz, 2 H, Ar), 7.26 - 7.36 (m, 6 H, Ar), 7.66 (d, *J*=8.3 Hz, 2 H, Ar), 7.80 - 7.93 (m, 6 H, Ar), 8.10 (d, *J*=8.9 Hz, 2 H, Ar), 8.57 (s, 1 H, -**CH**=N-).

$^{13}$C NMR: (75 MHz, DMSO-$d_6$) δ ppm: 25.23, 28.28, 28.45, 30.67, 34.64, 55.62, 67.65, 109.63, 114.25, 114.68, 118.89, 120.99, 121.80, 122.56, 126.90, 127.23, 128.71, 129.10, 130.47, 131.96, 132.76, 135.56, 143.14, 144.54, 148.43, 149.30, 160.03, 161.36, 163.70, 164.28.

Elemental analysis: Calculated for $C_{40}H_{36}N_2O_4$: C, 78.92 %, H, 5.96 %, N, 4.60 %. Found: C, 79.01 %, H, 6.00 %, N, 4.47 %.

***CB6OIBeO2*** *- 4-[{[4-({6-[4-(4-Cyanophenyl)phenyl]hexyl}oxy)phenyl]methylidene}amino] phenyl 4-ethoxybenzoate:*

Yield: 264 mg, 54.6 %.

Infrared $\bar{\nu}$ $cm^{-1}$: 3062 ($sp^2$ C-H); 2986, 2975, 2940, 2855 ($sp^3$ C-H); 2230 (C≡N); 1721 (C=O); 1603, 1571, 1512, 1495 (Ar C=C); 1256 (Ar C-O); 852, 837 (p. disubs. benzene C-H).

$^1$H NMR (300 MHz, DMSO-$d_6$) δ ppm: 1.31 - 1.55 (m, 7 H, -$CH_2$-$CH_2$-**$CH_2$**-**$CH_2$**-$CH_2$-$CH_2$-O-, -Ar-O-$CH_2$-**$CH_3$**), 1.64 (quin, *J*=7.5, 2 H, -$CH_2$-**$CH_2$**-$CH_2$-$CH_2$-$CH_2$-$CH_2$-O-), 1.75 (tt, *J*=7.0, 6.4 Hz, 2 H, -$CH_2$-$CH_2$-$CH_2$-$CH_2$-**$CH_2$**-$CH_2$-O-), 2.65 (t, *J*=7.5 Hz, 2 H, -Ar-**$CH_2$**-$CH_2$-), 4.05 (t, *J*=6.4 Hz, 2 H, -$CH_2$-**$CH_2$**-O-), 4.16 (q, J=7.0 Hz, -Ar-O-**$CH_2$**-$CH_3$), 7.06 (d, *J*=8.7 Hz, 2 H, Ar), 7.12 (d, *J*=9.0 Hz, 2 H, Ar), 7.25 - 7.37 (m, 6 H, Ar), 7.66 (d, *J*=8.3 Hz, 2 H, Ar), 7.82 - 7.94 (m, 6 H, Ar), 8.09 (d, *J*=9.0 Hz, 2 H, Ar), 8.57 (s, 1 H, -**CH**=N-).

Elemental analysis: Calculated for $C_{41}H_{38}N_2O_4$: C, 79.07 %, H, 6.15 %, N, 4.50 %. Found: C, 78.61 %, H, 6.22 %, N, 4.27 %.

***CB6OIBeO3*** *- 4-[{[4-({6-[4-(4-Cyanophenyl)phenyl]hexyl}oxy)phenyl]methylidene}amino] phenyl 4-propoxybenzoate:*

Yield: 283 mg, 57.4 %.

Infrared $\bar{\nu}$ $cm^{-1}$: 3051 ($sp^2$ C-H); 2969, 2940, 2906, 2867, 2853 ($sp^3$ C-H); 2223 (C≡N); 1722 (C=O); 1602, 1572, 1509, 1495 (Ar C=C); 1252 (Ar C-O); 845, 811 (p. disubs. benzene C-H). ).

$^1$H NMR (300 MHz, DMSO-$d_6$) δ ppm: 1.00 (t, J=7.4 Hz, 3H, -CH$_2$-**CH$_3$**), 1.32 - 1.56 (m, 4 H, -CH$_2$-CH$_2$-**CH$_2$**-**CH$_2$**-CH$_2$-CH$_2$-O-), 1.58 – 1.85 (m, 6 H, -CH$_2$-**CH$_2$**-CH$_2$-CH$_2$-**CH$_2$**-CH$_2$-O-, -O-CH$_2$-**CH$_2$**-CH$_3$), 2.65 (t, *J*=7.4 Hz, 2 H, -Ar-**CH$_2$**-CH$_2$-), 4.02 - 4.09 (two overlapping t, *J*=6.4 Hz, 4 H, -CH$_2$-CH$_2$-**CH$_2$**-O-, -O-**CH$_2$**-CH$_2$-CH$_3$), 7.06 (d, *J*=8.7 Hz, 2 H, Ar), 7.13 (d, *J*=9.0 Hz, 2 H, Ar), 7.25 - 7.39 (m, 6 H, Ar), 7.67 (d, *J*=8.3 Hz, 2 H, Ar), 7.82 - 7.95 (m, 6 H, Ar), 8.09 (d, *J*=8.9 Hz, 2 H, Ar), 8.57 (s, 1 H, -**CH**=N-).

Elemental analysis: Calculated for $C_{42}H_{40}N_2O_4$: C, 79.22 %, H, 6.33 %, N, 4.40 %. Found: C, 78.85 %, H, 6.35 %, N, 4.22 %.

***CB6OIBeO4*** *- 4-[{[4-({6-[4-(4-Cyanophenyl)phenyl]hexyl}oxy)phenyl]methylidene}amino] phenyl 4-butoxybenzoate:*

Yield: 301 mg, 59.1 %.

Infrared $\bar{\nu}$ $cm^{-1}$: 2935, 2868 ($sp^3$ C-H); 2225 (C≡N); 1723 (C=O); 1603, 1572, 1510, 1494 (Ar C=C); 1249 (Ar C-O); 837 (p. disubs. benzene C-H).

$^1$H NMR (300 MHz, DMSO-$d_6$) δ ppm: 0.95 (t, *J*=7.4 Hz, 3 H, -CH$_2$-**CH$_3$**), 1.30 - 1.55 (m, 6 H, -CH$_2$-CH$_2$-**CH$_2$**-**CH$_2$**-CH$_2$-CH$_2$-O-, -O-CH$_2$-CH$_2$-**CH$_2$**-CH$_3$), 1.62 (quin, J=7.4 Hz, 2 H, -Ar-CH$_2$-**CH$_2$**-CH$_2$-), 1.74 (tt, J=7.3, 6.4 Hz, 4 H, -CH$_2$-**CH$_2$**-CH$_2$-O-Ar-, -Ar-O-CH$_2$-**CH$_2$**-CH$_2$-), 2.65 (t, *J*=7.4 Hz, 2 H, -Ar-**CH$_2$**-CH$_2$-), 3.98 - 4.16 (Two overlapping t, J=6.4 Hz, 4 H, -CH$_2$-**CH$_2$**-O-Ar-, -Ar-O-**CH$_2$**-CH$_2$-), 7.05 (d, *J*=8.7 Hz, 2 H, Ar), 7.12 (d, *J*=8.9 Hz, 2 H, Ar), 7.23 - 7.40 (m, 6 H, Ar), 7.66 (d, *J*=8.1 Hz, 2 H, Ar), 7.80 - 7.94 (m, 6 H, Ar), 8.08 (d, *J*=8.9 Hz, 2 H, Ar), 8.57 (s, 1 H, -**CH**=N-).

Elemental analysis: Calculated for $C_{43}H_{42}N_2O_4$: C, 79.36 %, H, 6.51 %, N, 4.30 %. Found: C, 79.23 %, H, 6.56 %, N, 4.15 %.

***CB6OIBeO5*** *- 4-[{[4-({6-[4-(4-Cyanophenyl)phenyl]hexyl}oxy)phenyl]methylidene}amino] phenyl 4-pentoxybenzoate:*

Yield: 321 mg, 62.3 %.

Infrared $\bar{\nu}$ $cm^{-1}$: 2934, 2860 ($sp^3$ C-H); 2225 (C≡N); 1726 (C=O); 1604, 1573, 1511, 1494 (Ar C=C); 1248 (Ar C-O); 851, 835 (p. disubs. benzene C-H).

$^1$H NMR (300 MHz, DMSO-$d_6$) δ ppm: 0.91 (t, *J*=7.2 Hz, 3 H, -CH$_2$-**CH$_3$**), 1.28 - 1.55 (m, 8 H, -CH$_2$-CH$_2$-**CH$_2$**-**CH$_2$**-CH$_2$-CH$_2$-O-, -O-CH$_2$-CH$_2$-**CH$_2$**-**CH$_2$**-CH$_3$), 1.64 (tt, J=7.5, 7.2 Hz, 2 H, -Ar-CH$_2$-**CH$_2$**-CH$_2$-), 1.76 (tt, J=7.2, 6.5 Hz, -CH$_2$-**CH$_2$**-CH$_2$-O-Ar-, -Ar-O-CH$_2$-**CH$_2$**-CH$_2$-), 2.65 (t, *J*=7.5 Hz, 2 H, -Ar-**CH$_2$**-CH$_2$-), 3.98 - 4.16 (Two overlapping t, J=6.4 Hz, 4 H, -CH$_2$-**CH$_2$**-O-Ar-, -Ar-O-**CH$_2$**-CH$_2$-), 7.06 (d, *J*=8.9 Hz, 2 H, Ar), 7.12 (d, *J*=9.0 Hz, 2

H, Ar), 7.25 - 7.38 (m, 6 H, Ar), 7.67 (d, *J*=8.3 Hz, 2 H, Ar), 7.83 - 7.94 (m, 6 H, Ar), 8.08 (d, *J*=8.9 Hz, 2 H, Ar), 8.57 (s, 1 H, -**CH**=N-).
Elemental analysis: Calculated for $C_{44}H_{44}N_2O_4$: C, 79.49 %, H, 6.67 %, N, 4.21 %. Found: C, 79.47 %, H, 6.70 %, N, 4.08 %.

***CB6OIBeO6** - 4-[{[4-({6-[4-(4-Cyanophenyl)phenyl]hexyl}oxy)phenyl]methylidene}amino] phenyl 4-hexyloxybenzoate*:
Yield: 354 mg, 67.3 %.
Infrared $\bar{\nu}$ $cm^{-1}$: 2932, 2858 ($sp^3$ C-H); 2225 (C≡N); 1726 (C=O); 1604, 1573, 1511, 1494 (Ar C=C); 1247 (Ar C-O); 851, 835 (p. disubs. benzene C-H).
$^1$H NMR (300 MHz, DMSO-$d_6$) δ ppm: 0.89 (t, *J*=7.0 Hz, 3 H, -$CH_2$-**$CH_3$**), 1.27 - 1.52 (m, 10 H, -$CH_2$-$CH_2$-**$CH_2$**-**$CH_2$**-$CH_2$-$CH_2$-O-, -O-$CH_2$-$CH_2$-**$CH_2$**-**$CH_2$**-**$CH_2$**-$CH_3$), 1.64 (tt, *J*=7.3, 7.0 Hz, 2 H, -Ar-$CH_2$-**$CH_2$**-$CH_2$-), 1.75 (tt, *J*=7.0, 6.4 Hz, 4 H, -$CH_2$-**$CH_2$**-$CH_2$-O-Ar-, -Ar-O-$CH_2$-**$CH_2$**-$CH_2$-), 2.65 (t, *J*=7.4 Hz, 2 H, -Ar-**$CH_2$**-$CH_2$-), 4.01 - 4.13 (Two overlapping t, J=6.4 Hz, 4 H, -$CH_2$-**$CH_2$**-O-Ar-, -Ar-O-**$CH_2$**-$CH_2$-), 7.05 (d, *J*=8.7 Hz, 2 H, Ar), 7.11 (d, *J*=8.7 Hz, 2 H, Ar), 7.26 - 7.37 (m, 6 H, Ar), 7.66 (d, *J*=8.1 Hz, 2 H, Ar), 7.83 - 7.92 (m, 6 H, Ar), 8.08 (d, *J*=8.7 Hz, 2 H, Ar), 8.57 (s, 1 H, -**CH**=N-).
Elemental analysis: Calculated for $C_{45}H_{46}N_2O_4$: C, 79.62 %, H, 6.83 %, N, 4.13 %. Found: C, 79.63 %, H, 6.86 %, N, 3.98 %.

***CB6OIBeO7** - 4-[{[4-({6-[4-(4-Cyanophenyl)phenyl]hexyl}oxy)phenyl]methylidene}amino] phenyl 4-heptyloxybenzoate*:
Yield: 342 mg, 64.2 %.
Infrared $\bar{\nu}$ $cm^{-1}$: 3026 ($sp^2$ C-H); 2970, 2928, 2855 ($sp^3$ C-H); 2225 (C≡N); 1721 (C=O); 1603, 1572, 1511, 1494 (Ar C=C); 1253 (Ar C-O); 849, 826 (p. disubs. benzene C-H).
$^1$H NMR (300 MHz, DMSO-$d_6$) δ ppm: 0.88 (t, *J*=7.2 Hz, 3 H, -$CH_2$-**$CH_3$**), 1.23 - 1.55 (m, 12 H, -$CH_2$-$CH_2$-**$CH_2$**-**$CH_2$**-$CH_2$-$CH_2$-O-, -O-$CH_2$-$CH_2$-**$CH_2$**-**$CH_2$**-**$CH_2$**-**$CH_2$**-$CH_3$), 1.64 (tt, *J*=7.4, 7.2 Hz, 2 H, -Ar-$CH_2$-**$CH_2$**-$CH_2$-), 1.75 (tt, *J*=7.2, 6.4 Hz, 4 H, -$CH_2$-**$CH_2$**-$CH_2$-O-Ar-, -Ar-O-$CH_2$-**$CH_2$**-$CH_2$-), 2.65 (t, *J*=7.4 Hz, 2 H, -Ar-**$CH_2$**-$CH_2$-), 4.00 - 4.13 (Two overlapping t, J=6.4 Hz, 4 H, -$CH_2$-**$CH_2$**-O-Ar-, -Ar-O-**$CH_2$**-$CH_2$-), 7.06 (d, *J*=8.9 Hz, 2 H, Ar), 7.12 (d, *J*=9.0 Hz, 2 H, Ar), 7.24 - 7.38 (m, 6 H, Ar), 7.66 (d, *J*=8.3 Hz, 2 H, Ar), 7.82 - 7.94 (m, 6 H, Ar), 8.08 (d, *J*=8.9 Hz, 2 H, Ar), 8.57 (s, 1 H, -**CH**=N-).
Elemental analysis: Calculated for $C_{46}H_{48}N_2O_4$: C, 79.74 %, H, 6.98 %, N, 4.04 %. Found: C, 79.70 %, H, 6.95 %, N, 3.93 %.

***CB6OIBeO8** - 4-[{[4-({6-[4-(4-Cyanophenyl)phenyl]hexyl}oxy)phenyl]methylidene}amino] phenyl 4-octyloxybenzoate*:
Yield: 339 mg, 62.0 %.
Infrared $\bar{\nu}$ $cm^{-1}$: 2927, 2852 ($sp^3$ C-H); 2226 (C≡N); 1721 (C=O); 1604, 1572, 1512, 1494 (Ar C=C); 1252 (Ar C-O); 851, 827 (p. disubs. benzene C-H).
$^1$H NMR (300 MHz, DMSO-$d_6$) δ ppm: 0.87 (t, *J*=7.0 Hz, 3 H, -$CH_2$-**$CH_3$**), 1.22 - 1.55 (m, 14 H, -$CH_2$-$CH_2$-**$CH_2$**-**$CH_2$**-$CH_2$-$CH_2$-O-, -O-$CH_2$-$CH_2$-**$CH_2$**-**$CH_2$**-**$CH_2$**-**$CH_2$**-**$CH_2$**-$CH_3$), 1.64 (tt, *J*=7.6, 7.2 Hz, 2 H, -Ar-$CH_2$-**$CH_2$**-$CH_2$-), 1.75 (tt, *J*=7.2, 6.4 Hz, 4 H, -$CH_2$-**$CH_2$**-$CH_2$-O-Ar-, -Ar-O-$CH_2$-**$CH_2$**-$CH_2$-), 2.65 (t, *J*=7.6 Hz, 2 H, -Ar-**$CH_2$**-$CH_2$-), 4.01 - 4.13 (Two overlapping t, J=6.4 Hz, 4 H, -$CH_2$-**$CH_2$**-O-Ar-, -Ar-O-**$CH_2$**-$CH_2$-), 7.06

(d, *J*=8.9 Hz, 2 H, Ar), 7.12 (d, *J*=9.0 Hz, 2 H, Ar), 7.27 - 7.37 (m, 6 H, Ar), 7.67 (d, *J*=8.3 Hz, 2 H, Ar), 7.84 - 7.93 (m, 6 H, Ar), 8.08 (d, *J*=8.9 Hz, 2 H, Ar), 8.58 (s, 1 H, -**CH**=N-).
Elemental analysis: Calculated for $C_{47}H_{50}N_2O_4$: C, 79.86 %, H, 7.13 %, N, 3.96 %. Found: C, 79.96 %, H, 7.18 %, N, 3.81 %.

***CB6OIBeO9*** *- 4-[{[4-({6-[4-(4-Cyanophenyl)phenyl]hexyl}oxy)phenyl]methylidene}amino] phenyl 4-nonyloxybenzoate*:
Yield: 312 mg, 55.1 %.
Infrared $\bar{\nu}$ $cm^{-1}$: 2922, 2868, 2851 ($sp^3$ C-H); 2226 (C≡N); 1717 (C=O); 1602, 1572, 1507, 1495 (Ar C=C); 1246 (Ar C-O); 838, 821 (p. disubs. benzene C-H).
$^1$H NMR (300 MHz, DMSO-$d_6$) δ ppm: 0.87 (t, *J*=7.1 Hz, 3 H, -$CH_2$-**$CH_3$**), 1.22 - 1.51 (m, 16H, -$CH_2$-$CH_2$-**$CH_2$**-**$CH_2$**-$CH_2$-$CH_2$-O-, -O-$CH_2$-$CH_2$-**$CH_2$**-**$CH_2$**-**$CH_2$**-**$CH_2$**-**$CH_2$**-**$CH_2$**-$CH_3$), 1.64 (tt, *J*=7.4, 7.2 Hz, 2 H, -Ar-$CH_2$-**$CH_2$**-$CH_2$-), 1.75 (tt, *J*=7.2, 6.4 Hz, 4 H, -$CH_2$-**$CH_2$**-$CH_2$-O-Ar-, -Ar-O-$CH_2$-**$CH_2$**-$CH_2$-), 2.65 (t, *J*=7.4 Hz, 2 H, -Ar-**$CH_2$**-$CH_2$-), 4.00 - 4.14 (Two overlapping t, J=6.4 Hz, 4 H, -$CH_2$-**$CH_2$**-O-Ar-, -Ar-O-**$CH_2$**-$CH_2$-), 7.07 (d, *J*=8.9 Hz, 2 H, Ar), 7.12 (d, *J*=8.9 Hz, 2 H, Ar), 7.24 - 7.40 (m, 6 H, Ar), 7.67 (d, *J*=8.5 Hz, 2 H, Ar), 7.82 - 7.96 (m, 6 H, Ar), 8.08 (d, *J*=8.9 Hz, 2 H, Ar), 8.58 (s, 1 H, -**CH**=N-).
Elemental analysis: Calculated for $C_{48}H_{52}N_2O_4$: C, 79.97 %, H, 7.27 %, N, 3.89 %. Found: C, 79.98 %, H, 7.31 %, N, 3.73 %

***CB6OIBeO10*** *- 4-[{[4-({6-[4-(4-Cyanophenyl)phenyl]hexyl}oxy)phenyl]methylidene}amino] phenyl 4-heptyloxybenzoate*:
Yield: 304 mg, 53.6 %.
Infrared $\bar{\nu}$ $cm^{-1}$: 3036 ($sp^2$ C-H); 2936, 2917, 2848 ($sp^3$ C-H); 2225 (C≡N); 1718 (C=O); 1603, 1572, 1510, 1495 (Ar C=C); 1251 (Ar C-O); 844, 811 (p. disubs. benzene C-H).
$^1$H NMR (300 MHz, DMSO-$d_6$) δ ppm: 0.86 (t, *J*=7.0 Hz, 3 H, -$CH_2$-**$CH_3$**), 1.08 - 1.71 (m, 20H, -$CH_2$-**$CH_2$**-**$CH_2$**-**$CH_2$**-$CH_2$-$CH_2$-O-, -O-$CH_2$-$CH_2$-**$CH_2$**-**$CH_2$**-**$CH_2$**-**$CH_2$**-**$CH_2$**-**$CH_2$**-**$CH_2$**-$CH_3$), 1.75 (tt, *J*=7.2, 6.4 Hz, 4 H, -$CH_2$-**$CH_2$**-$CH_2$-O-Ar-, -Ar-O-$CH_2$-**$CH_2$**-$CH_2$-), 2.66 (t, *J*=7.0 Hz, 2 H, -Ar-**$CH_2$**-$CH_2$-), 4.01 - 4.13 (Two overlapping t, J=6.4 Hz, 4 H, -$CH_2$-**$CH_2$**-O-Ar-, -Ar-O-**$CH_2$**-$CH_2$-), 7.06 (d, *J*=8.9 Hz, 2 H, Ar), 7.12 (d, *J*=9.0 Hz, 2 H, Ar), 7.26 - 7.38 (m, 6 H, Ar), 7.67 (d, *J*=8.3 Hz, 2 H, Ar), 7.84 - 7.93 (m, 6 H, Ar), 8.08 (d, *J*=8.7 Hz, 2 H, Ar), 8.58 (s, 1 H, -**CH**=N-).
Elemental analysis: Calculated for $C_{49}H_{54}N_2O_4$: C, 80.08 %, H, 7.41 %, N, 3.81 %. Found: C, 80.25 %, H, 7.45 %, N, 3.66 %

## Experimental methods

Calorimetric studies were performed with TA DSC Q200 calorimeter, samples of mass 1-3 mg were sealed in aluminum pans and kept in nitrogen atmosphere during measurement, both heating and cooling scans were performed with rate 10 K/min.
Wide angle X-ray diffractograms were obtained with Bruker D8 GADDS system (CuKα line, Goebel mirror, point beam collimator, Vantec2000 area detector ). Samples were prepared as droplets on heated surface. The temperature dependence of layer thickness was determined from small-angle X-ray diffraction experiments performed with Bruker D8 Discover system (CuKα line, Goebel mirror, Anton Paar DCS350 heating stage, scintillation counter) working in reflection mode, homeotropically aligned samples were prepared as thin film on a silicon wafer. The tilt angle in tilted smectic phase was calculated from decrease of layer spacing, $d_{SmC} = d_{SmA} \cos(\theta)$, with respect layer thickness value extrapolated from the data recorded in orthogonal SmA phase to temperature range of the tilted phase.

The resonant x-ray experiment was performed on the soft x-ray scattering beam line (11.0.1.2) at the Advanced Light Source of Lawrence Berkeley National Laboratory. The x-ray beam was tuned to the K-edge of carbon strongest absorption peak, ~283.6 eV, however for the studied dimeric material few more peaks with different energy were also detected. An energy scan was done at two temperatures: 392 and 360 K in $SmC_{TB}$ and HexI phase, respectively. The x-ray beam with a cross-section of 300 × 200 μm was linearly polarized, with the polarization direction that can be continuously changed from the horizontal to vertical. Samples with thickness lower than 1 $\mu$m were placed between two 100-nm-thick $Si_3N_4$ slides. The scattering intensity was recorded using the Princeton PI-MTE CCD detector, cooled to $-45°C$, having a pixel size of 27 $\mu$m, with an adjustable distance from the sample. The detector was translated off axis to enable a recording of the diffracted x-ray intensity. The adjustable position of the detector allowed to cover a broad range of $q$ vectors, corresponding to periodicities from approximately 5.0 – 300 nm.

Optical studies were performed using the Zeiss Imager A2m polarizing microscope equipped with Linkam heating stage. Samples were observed in glass cells with various thickness: 1.8 to 10 microns. The microscope setup was equipped with Abrio system for precise birefringence measurements. The birefringence was calculated from optical retardation at green light (λ=546nm). The average retardation was measured in 10x10 micron region, but space modulation of retardation was also evaluated with submicron resolution. For birefringence measurements the 3 micron cells were used with planar alignment layer. Since the system allows to measure retardation only up to 273 nm therefore the absolute value of retardation was determined by comparing the results obtained for the samples with different thickness, 1.8 and 5 micron.

The conical tilt angle in the $SmC_{TB}$ phase was deduces form decrease of birefringence in respect to those measured in SmA phase, $\Delta n_{SmCTB}=\Delta n_{SmA}(3\cos^2(\theta)-1)/2$, [6], the birefringence of SmA phase was extrapolated to the lower temperatures range, assuming linear temperature dependence of Δn in SmA, the increase of birefringence in SmA phase with lowering temperature is due to the growing positional and/or conformational order of molecules.

AFM images have been taken with Bruker Dimension Icon microscope, working in tapping mode at liquid crystalline-air surface. Cantilevers with a low spring constant, $k = 0.4\ Nm^{-1}$ were used, the resonant frequency was in a range of 70-80 kHz, typical scan frequency was 1 Hz.

The CD spectra were collected with a J-815 spectropolarimeter from Jasco, Japan, using quartz plates as substrates for samples.

**Additional results**

XRD studies

Both nematic phases, N and $N_{TB}$, show similar x-ray patterns (Fig. S1), in the low angle region very weak diffraction signals were visible, corresponding to full (L) and half molecular length (L/2). However, upon transition to $N_{TB}$ phase the partial alignment of the sample present in N phase is lost.

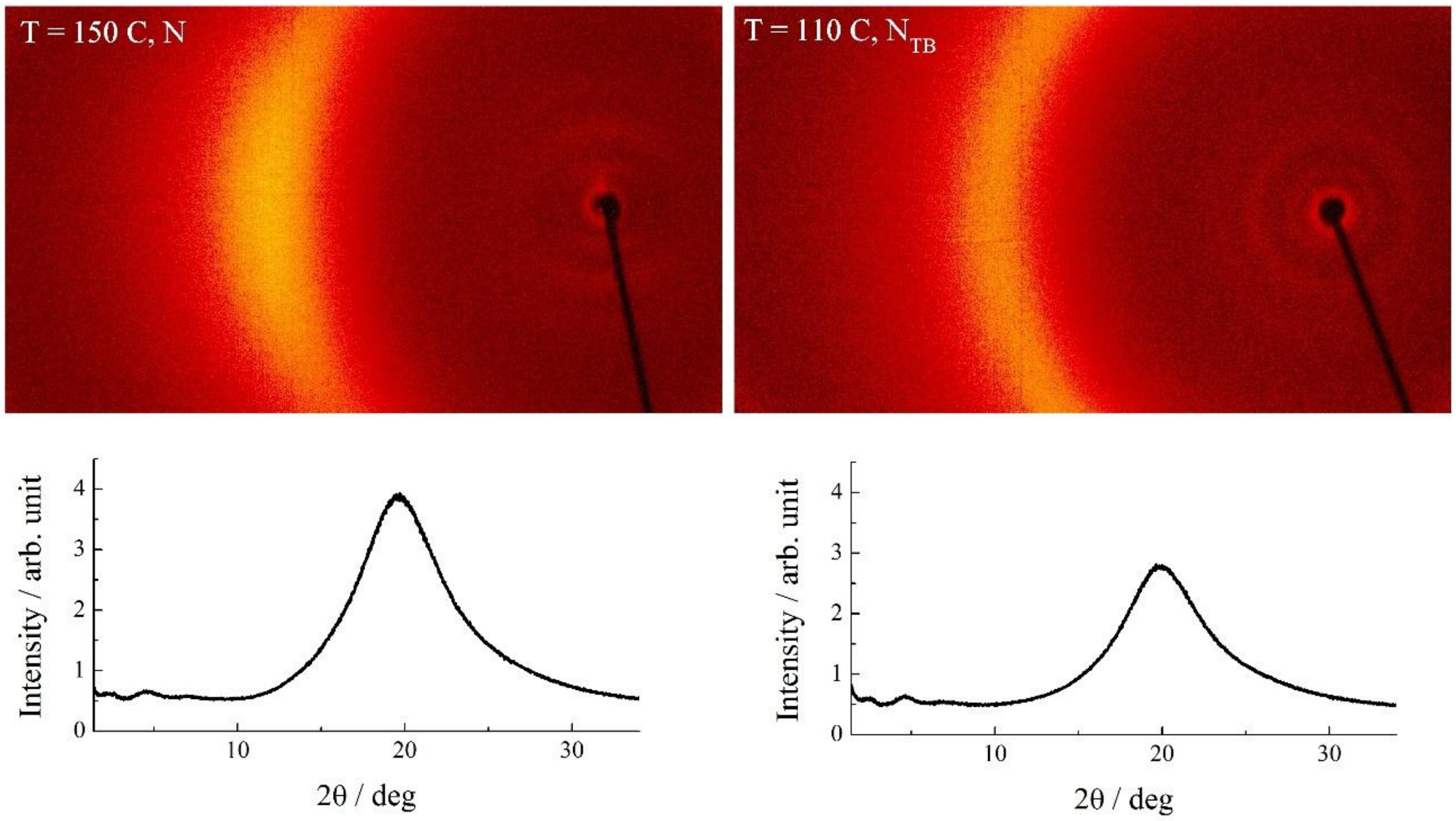

Figure S1. Wide angle 2D XRD patterns and corresponding dependence of diffracted intensity vs. diffraction angle obtained for nematic (N) and twist-bend nematic ($N_{TB}$) phases of homologue n=5.

Lowering temperature, for homologues with n>6 exhibiting SmA phase below N phase, continuously decreases the width of the low-angle signal corresponding to full molecular length, showing that system has tendency to lock to lamellar structure with d=L. X-ray diffraction measurements performed for compound n=8, with a sequence of 4 smectic phases below nematic phase, revealed that three phases, assigned as: SmA, $SmA_b$ and $SmC_{TB}$, have no long-range positional order within the smectic layers (liquid-like smectic phases). The transition to the lowest temperature lamellar phase, HexI, is marked by sudden change in the high angle range of the x-ray pattern, the diffused signal observed in SmA, $SmA_b$, $SmC_{TB}$ phases, narrows and becomes doubled (Fig. S2). The equatorial position of one of the high angle signals, indexed as (020), unambiguously determines the tilt direction toward nearest neighbour molecules in the in-plane hexagonal lattice, thus identifies HexI phase.

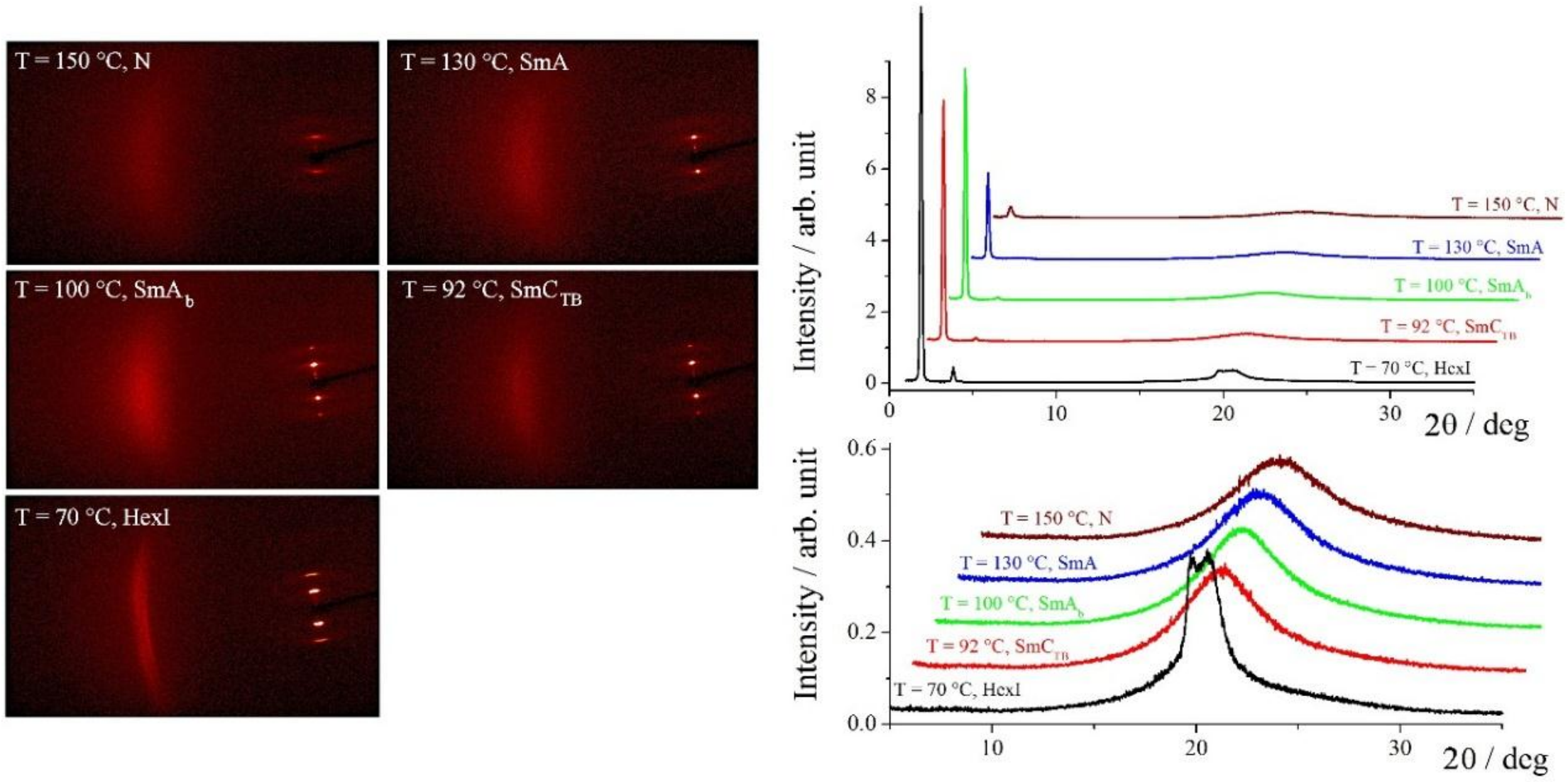

Figure S2. (left) Wide angle 2D XRD patterns obtained for aligned sample of compound n=8. (right) Integrated intensity of diffraction signals vs. diffraction angle 2θ. Lower graph show magnified high angle range of the diffractograms.

The evolution of the layer spacing at SmA-SmC$_{TB}$ phase transition points toward the weakly first order transition for homologue n=7, as evidenced by slight increase of the width of the diffraction signal caused by two phase co-existence in ~0.5K temperature range close to the transition temperature, and continuous transition for homologue n=8 (Fig. S3). The tilt angle calculated from the layer thickness decreases, following critical behaviour $\theta=\theta_0((T_c-T)/T_c)^{\beta}$ with critical exponent β~0.15 and 0.40 for compounds n=7 and n=8, respectively.

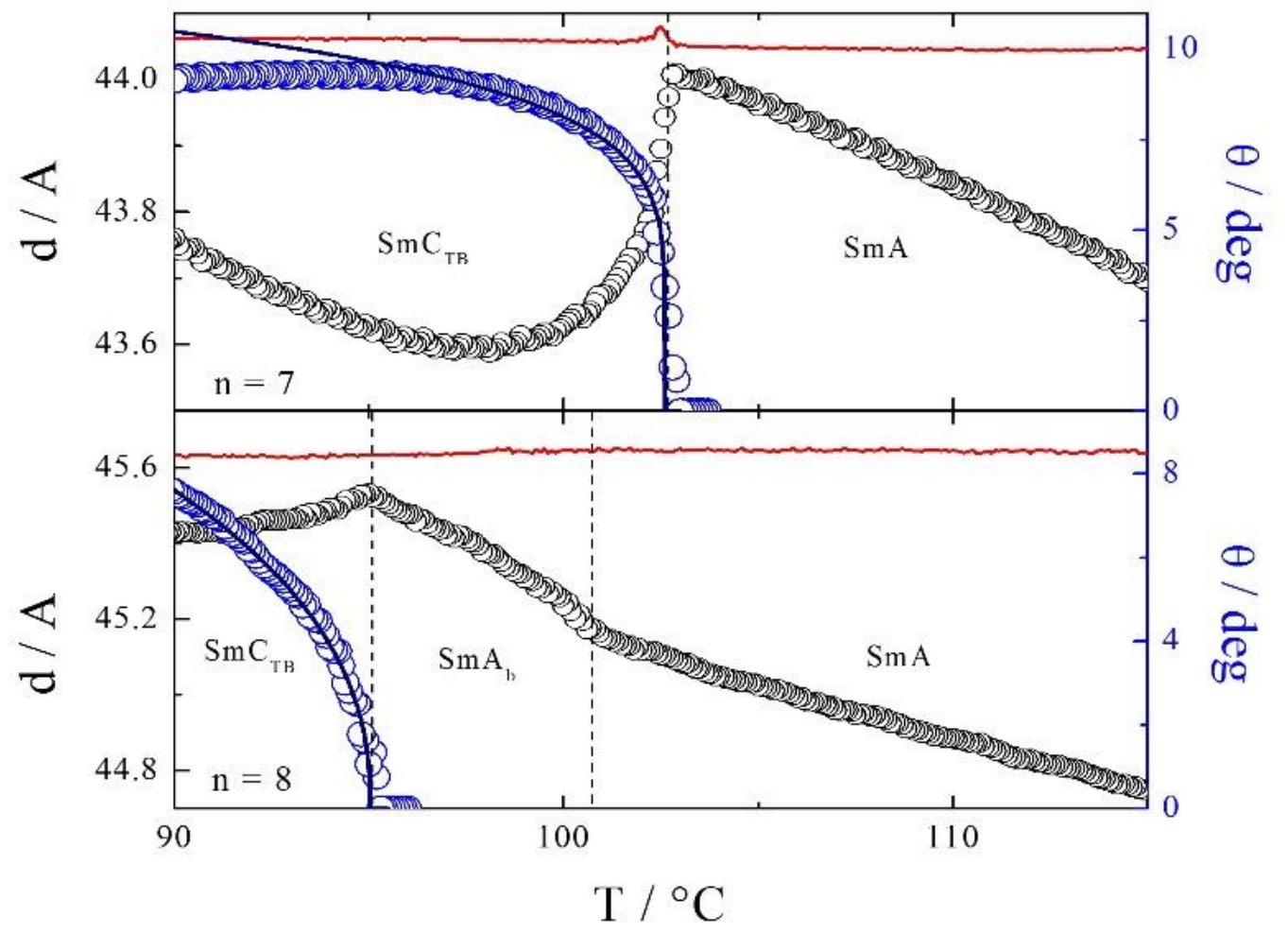

Figure S3. Temperature evolution of layer thickness (black circles) near the transition between orthogonal and tilted smectic phases for homologues n=7 and n=8. For homologue n=7 the SmA-$SmC_{TB}$ transition is weakly of first order, as evidenced by appearance of two-phase region manifesting as slight increase of the width of the diffraction signal related to layer periodicity (red line), the $SmA_b$-$SmC_{TB}$ phase transition in homologue n=8 is continuous. Temperature dependence of tilt angle (blue circles), obtained from the decrease of layer spacing in $SmC_{TB}$ phase with respect layer thickness value extrapolated from the data recorded in orthogonal phase, has been fitted to power law with critical exponents β equal 0.15 and 0.40 for compounds n=7 and n=8, respectively.

The resonant x-ray diffraction studies clearly revealed the signal from superstructure in HexI and $SmC_{TB}$ phases (Fig S4). The small temperature range (2-3 K) in which both signals coexists points to first order transition.

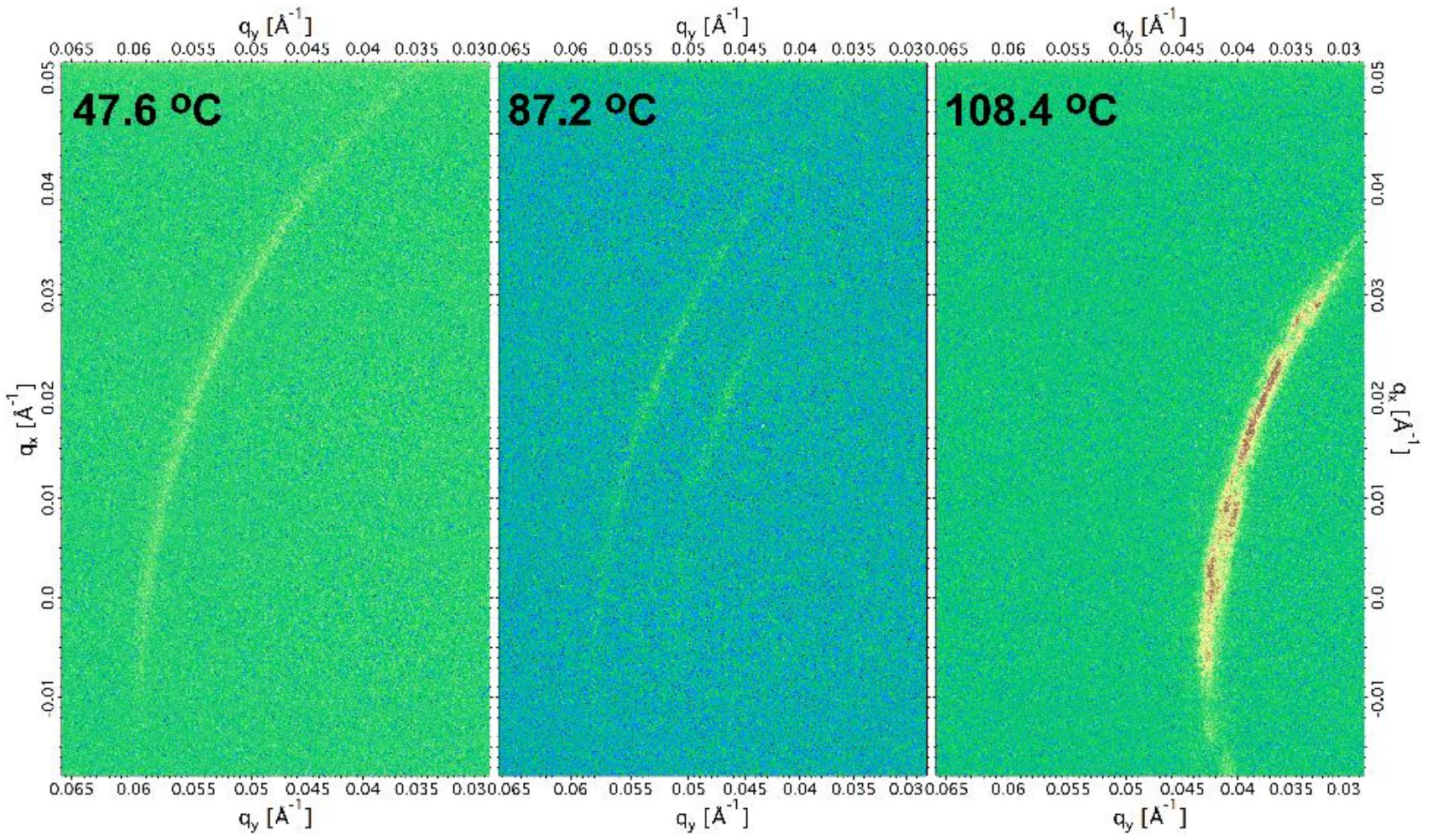

Figure S4. 2D RSoXS pattern measured for homologue n=7 in the temperature range of the HexI (47.6 °C) $SmC_{TB}$ (108.4 °C) and in the phase coexistence range (87.2 °C).

Optical and AFM studies

In both $N_{TB}$ and $SmC_{TB}$, phases, in the cells with planar anchoring the characteristic stripe texture was found, the periodicity of the stripes was defined by the cell thickness (Fig. S5). Moreover, by studying the mixture of homologues n=6 and n=7 (30 wt. % of homologue n=7) it was confirmed that the stripe pattern is not affected by the $N_{TB}$ - $SmC_{TB}$ phase transition.

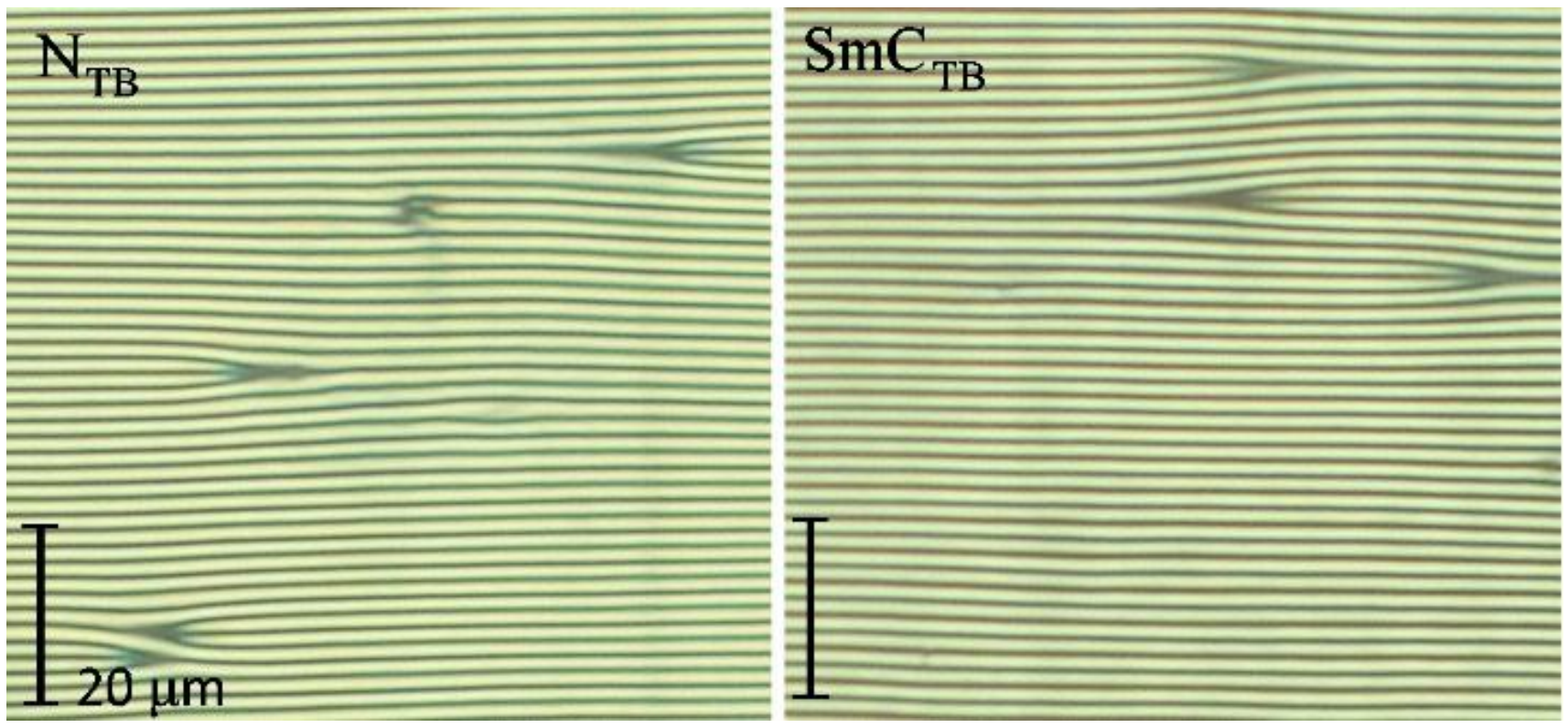

Figure S5. Stripe texture observed in $N_{TB}$ (homologue n=6) and $SmC_{TB}$ (homologue n=7) phases in 3 micron cells with planar anchoring.

The optical texture of HexI phase, in the cell with homeotropic anchoring, observed with slightly de-crossed polarizers revealed presence of domains with small optical activity (Fig. S6). The source of optical activity is unclear, it could be due to the lowering of the local symmetry by molecular tilt combined with local biaxial order (so called 'layer chirality' [7]) or due to formation of chiral excimers made of twisted molecular pairs [8].

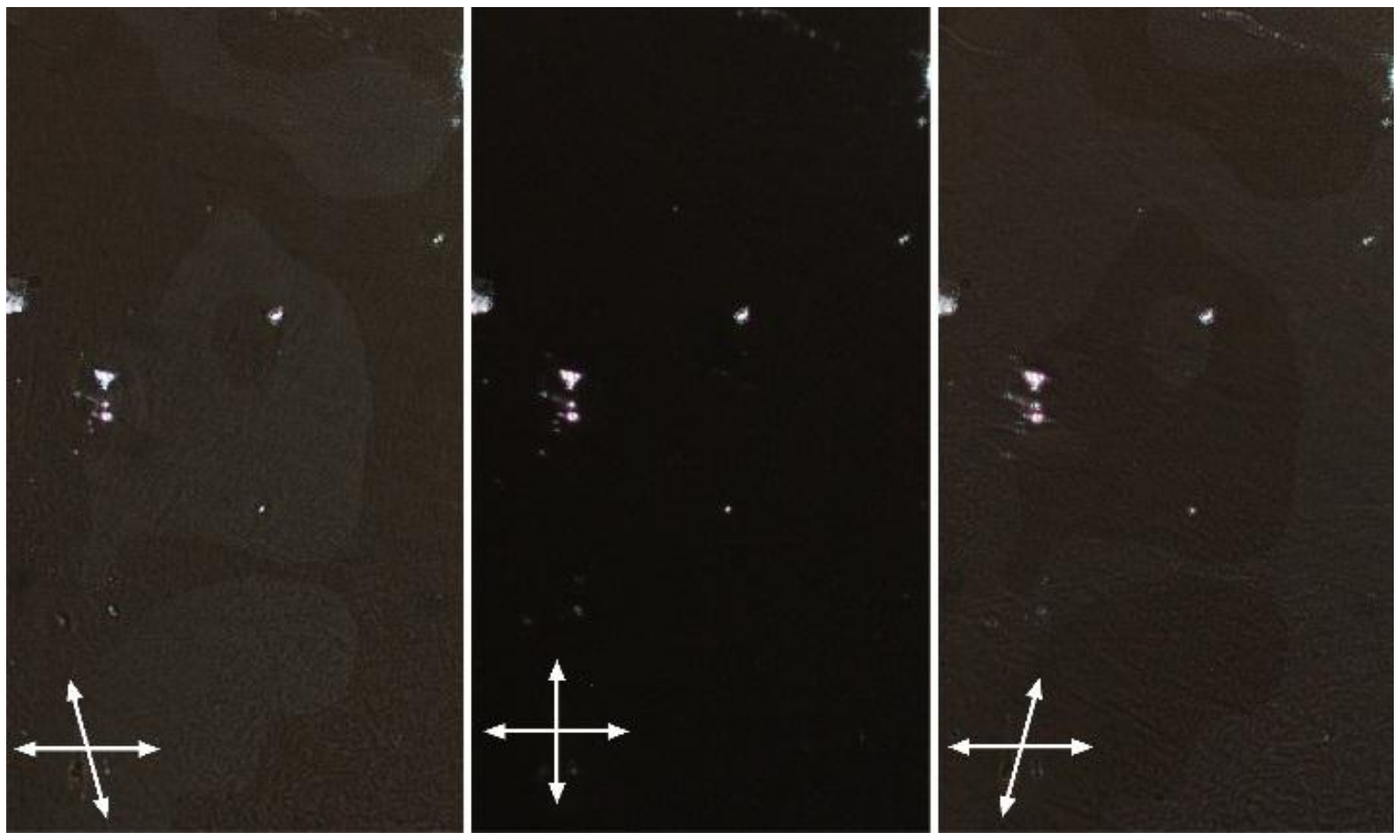

Figure S6. Domains with optical activity observed in HexI phase of homologue n=7 by slight de-crossing of polarizers.

The chiral nature of the HexI phase is consistent with observation of CD signal, at the wavelength of optical absorption band of material, 350nm (Fig. S7); the absorption at this wavelength is due to the presence of imine unit in molecular structure. On heating, the CD signal decreases and disappears at the transition to $SmC_{TB}$ (n=7), to SmA phase (n=10) or to $N_{TB}$ phase (n=6). Lack of CD signal in $N_{TB}$ and $SmC_{TB}$ phases, indicates that chiral domains in these phases are smaller than the wavelength of visible light, which averages the chiral effects to zero. Interestingly, for material n=7 in some heating runs the CD signals reverses its sign upon the sample re-crystallization (HexI phase in this material is metastable). The non-zero CD signal suggests that also crystalline phase has chiral structure, however this should be taken with some caution as the CD data can be unambiguously analysed only for non birefringent samples.

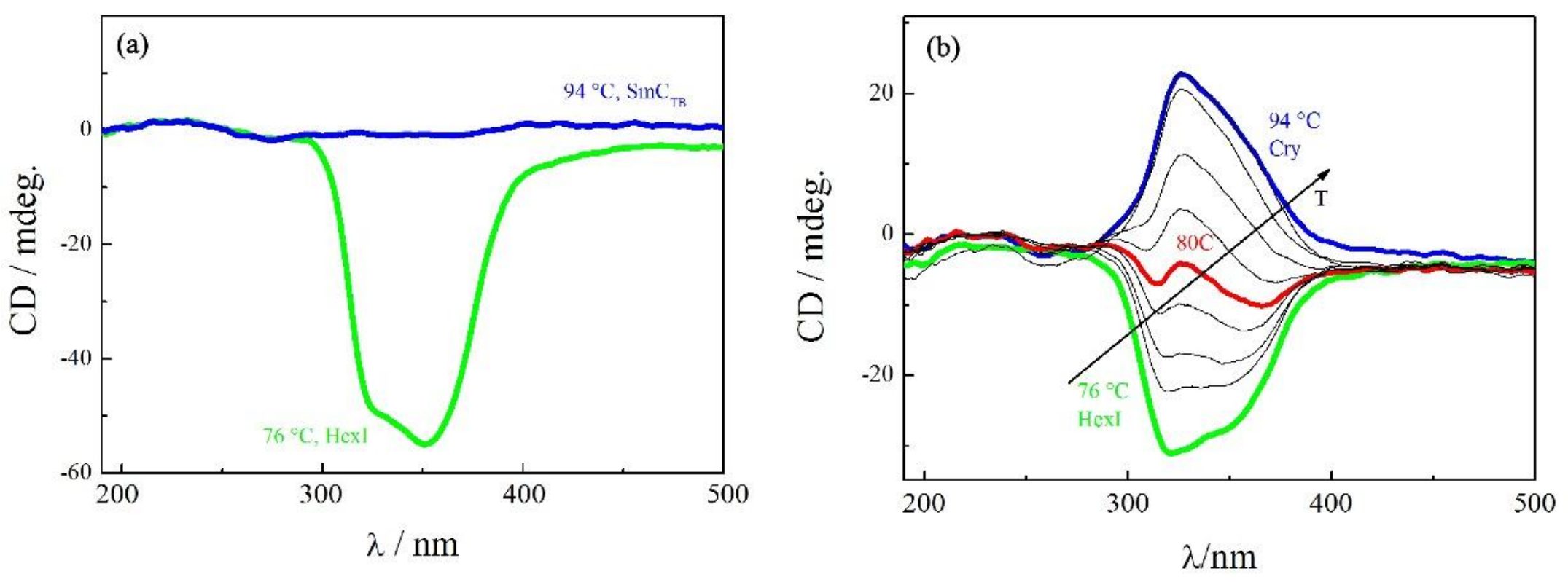

Figure S7. Evolution of CD signal for homologue n=7 on heating from HexI to $SmC_{TB}$ phase (a) and upon the re-crystallization of the sample heated from HexI phase (b).

The entangled helical filament morphology was observed at room temperature for quickly cooled samples (the procedure preserved HexI phase) by AFM method (Fig. S8a). Upon sample recrystallization the morphology changed and the plate like domains are formed (Fig. S8b).

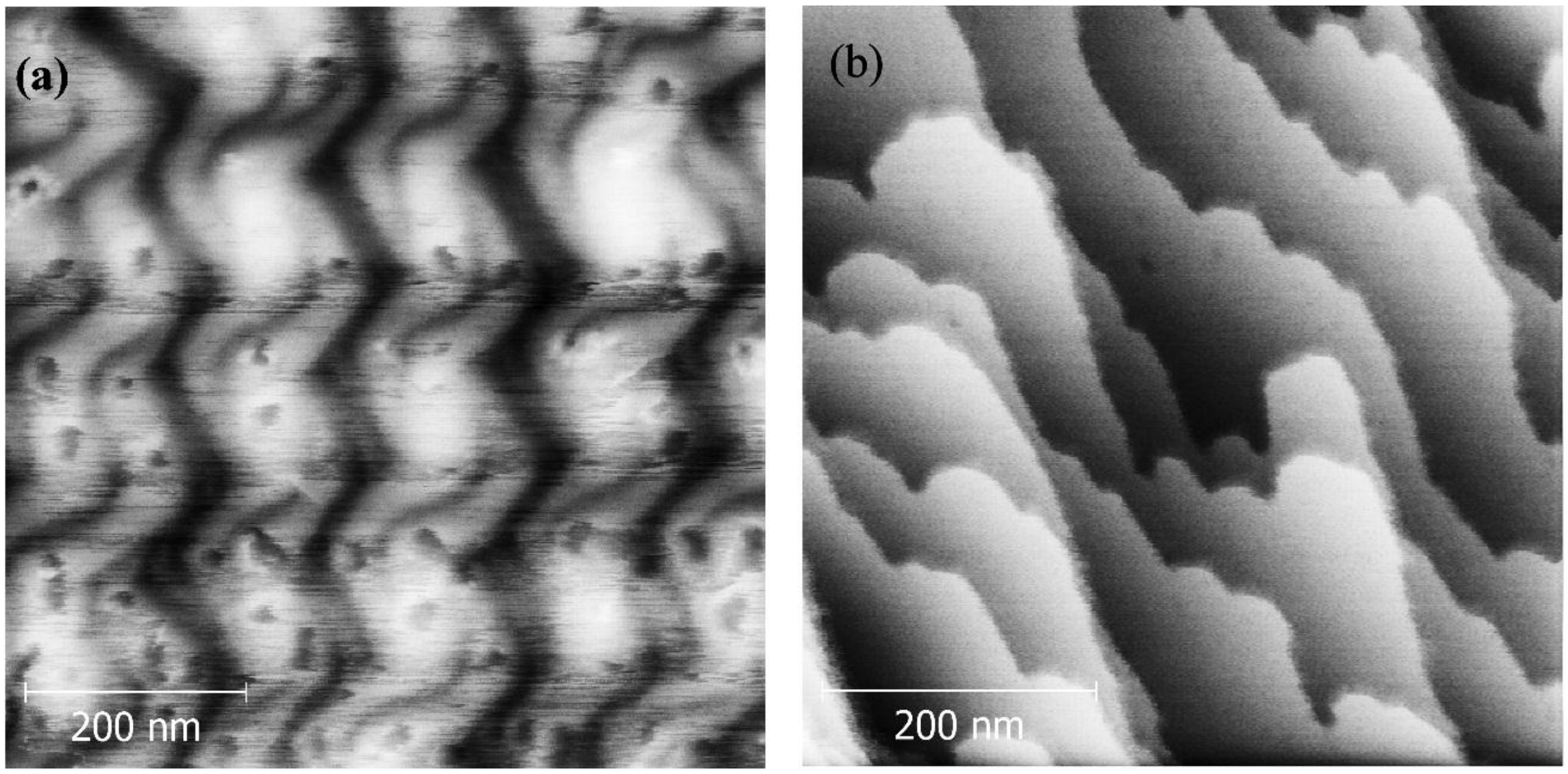

Figure S8. (a) AFM image of filaments formed by HexI phase observed at room temperature. Re-crystallization of the sample is accompanied by the change of morphology to the flat plates (b). Images were taken for homologues n=7 (a) and n=5 (b) at room temperature.